\documentclass[a4paper,11pt]{article}
\usepackage[utf8]{inputenc}
\usepackage
 {
   amsmath,amssymb,graphicx,tabularx
 }
\usepackage{color}
\usepackage{hyperref}
 \usepackage{cleveref}
\usepackage{caption,subcaption,cite}
\usepackage{geometry}
\usepackage{authblk}
\usepackage{fontenc}
\usepackage[titletoc,title]{appendix}
\usepackage{cancel}
 \usepackage[dvipsnames]{xcolor}
{\em }
\definecolor{darkgreen}{cmyk}{1,0,1,0.4}

\title{Distinguishing CP-odd couplings of the Higgs boson to weak boson pairs}
\author{ Siddharth Dwivedi$^1$\footnote{E-mail: siddharthdwivedi@hri.res.in}, Dilip Kumar Ghosh$^2$\footnote{E-mail: tpdkg@iacs.res.in},
         Biswarup Mukhopadhyaya$^1$\footnote{E-mail: biswarup@hri.res.in} 
        and Ambresh Shivaji$^3$\footnote{E-mail: ambresh.shivaji@pv.infn.it}}
  \affil{$^1$\textit{Regional Centre for Accelerator-based Particle Physics,}
  \textit{Harish-Chandra Research Institute, Chhatnag Road, Jhunsi,} \\
  \textit{Allahabad - 211019, India}}
 \affil{$^2$\textit{Department of Theoretical Physics,}
\\
 \textit{Indian Association for the Cultivation of Science,}
\\
 \textit{2A \& 2B Raja S.C. Mullick Road,}\\

 \textit{Kolkata - 700032, India}}
 \affil{$^3$\textit{INFN, Sezione di Pavia,} \\
  \textit{Via A. Bassi 6, 27100 Pavia, Italy}}

\begin{document}

\maketitle

\begin{abstract}

We consider the observable effects of CP-violating anomalous 
 $ZZh$ interaction arising from gauge invariant dimension-6 operators
at the Large Hadron Collider (LHC), with the purpose of distinguishing
them from not only the standard model effects but also those of  CP-even
anomalous interactions of similar nature. The postulation of a gauge 
invariant origin makes various couplings of this kind interrelated.
The updated constraints from the LHC as well as limits from neutron 
and electron dipole moments are used in selecting the benchmark 
interaction strengths. We use some asymmetry parameters that have no
contribution from standard or CP-even anomalous interactions. Parton showering
and detector level simulation is included to make our analysis as 
realistic as possible. On the whole, we conclude that gauge invariant interaction
of strength $\geq 40/\rm TeV^{2} $ can be successfully isolated using integrated
luminosities in the  1.5-3.0 $\rm ab^{-1}$ range.
\end{abstract}

\section{Introduction}

As results from the high energy run of the Large Hadron Collider (LHC)
start arriving, interest in probing any non-standard behavior of the
scalar with mass around 125-126 GeV  \cite{Aad:2012tfa, Chatrchyan:2012ufa} remains
ever-alive.  To keep the flame on, departure from the pattern of its
interactions as laid down in the standard model (SM) is still allowed,
though not imperatively, by the (7+8) TeV data. On the other hand,
issues ranging from the naturalness of the Higgs mass to the question
of vacuum stability continue to nudge theorists even after Higgs
discovery.  Therefore, it is both legitimate and well-motivated to
enquire if the high energy run can reveal some non-standard
interactions of the particle which, in all fairness, can be called `a Higgs boson'. 
Some such investigations in relation to CP-violating $ZZh$
coupling are reported here.

A rather general approach is to consider effective
interactions of the Higgs, without committing oneself to any specific
theoretical model. In this spirit, many studies have taken place
over the years, probing higher-dimensional $WWh$ and $ZZh$
interactions, often taking them in isolation and predicting observable
phenomena in terms of the new operators
\cite{Gao:2010qx,Choi:2002jk,DeRujula:2010ys,Boughezal:2012tz,Stolarski:2012ps,Djouadi:2013yb,Godbole:2013lna,Ellis:2013ywa,Sun:2013yra,Anderson:2013afp,
  Delaunay:2013npa,Maltoni:2013sma,Gavela:2014vra,Banerjee:2013apa,Amar:2014fpa,Ellis:2014dva,Desai:2011yj,Bolognesi:2012mm,Chen:2014ona,CMS_report_h_4l_2014,
  Khachatryan:2014kca,CMS_report_WWh_2014,Plehn:2001nj,Hankele:2006ma,Miller:2001bi,Han:2000mi,Biswal:2008tg,Hagiwara:2000tk,Chang:1993jy,Biswal:2009ar,
  Dutta:2008bh,Biswal:2012mp,Han:2009ra,Christensen:2010pf,Englert:2012xt,Biswal:2005fh,Godbole:2007cn,Bhattacherjee:2015xra,Beneke:2014sba,Godbole:2014cfa,
  Dawson:2013owa,Kruse:2014pya,Rattazzi:1988ye,Hagiwara:1993sw,Gounaris:1995mx,Niezurawski:2004ga,Skjold:1995jp,Skjold:1993jd,Skjold:1994qn,Grzadkowski:1995rx,
  Godbole:2006uy,Zhang:2003it,Buszello:2006hf,Hankele:2006ja}.  It is,
however, worthwhile to obtain these interactions in terms of {\em
  gauge invariant higher-dimensional operators}
\cite{Buchmuller:1985jz,Grzadkowski:2010es,Contino:2013kra,Leung:1984ni,Hagiwara:1993qt,GonzalezGarcia:1999fq}.
This is primarily because these operators can arise, if at all, by
integrating out hitherto unseen new physics, and such physics,
ruling above the electroweak symmetry breaking scale, must by
default be $SU(2)_L \times U(1)_Y$ invariant. Also, the $WWh, ZZh,
\gamma \gamma h$ and $\gamma Z h$ interaction terms ensuing from such
gauge invariant operators are correlated, and this correlation
percolates into observed phenomena involving Higgs production as well
as decay. A detailed inventory as well as a large mass of work already
exists on CP-even operators of this kind and their constraints and
consequences. Systematic studies on their CP-odd counterparts are
relatively sparse, and perhaps less exhaustive. However, the very
observation of the baryon asymmetry in the universe strongly suggests
some source of CP-violation over and above what is manifested in the
Cabibbo-Kobayashi-Maskawa scheme of the SM. A bottom-up investigation
of CP-odd couplings of the recently discovered scalar is therefore
as well-founded as one on their CP-even counterparts.

In general, there are five each of CP-even and-odd operators of
dimension-6 involving $VVh$ interactions. 
Coefficients of the latter are subject to rather strong
constraints from the limits on the electric dipole moments (EDM) of
the neutron and the electron and also from the upper bounds on Hg EDM 
\cite{Chien:2015xha, Cirigliano:2016njn}, in addition to global fits of collider
data (including those from the LHC) that constrain the CP-even
operators as well. Arguably a detailed study of these
constraints on the CP-violating operators was presented in the
reference \cite{Dwivedi:2015nta}. It is, however, possible to
generalize the study even further. First, the constraints from EDM are
rather tight on individual operators which tend to get very strongly
correlated if there are two or three of them  present
at a time. On the other
hand, taking them together along with other dimension-6 operators 
constructed out of the fermion fields and the Higgs doublet
\cite{Buchmuller:1985jz, Grzadkowski:2010es}, which can potentially
contribute to fermion EDMs, can relax such correlations
considerably, and one can end up with large values of the effective
$VVh$ couplings that potentially contribute to collider observables.
Moreover, once we introduce the gauge invariant effective interaction terms,
the simultaneous existence of several of them is the general expectation,
unless some specific symmetries are invoked to forbid some of them.
Secondly, global fits of collider data can yield less stringent
constraints if CP-odd and-even operators are all present
simultaneously, mainly due the sheer proliferation of parameters
(which nonetheless should be there in the general effective theory).
On the other hand, it is only natural to expect operators of both
kinds of a given dimensionality to exist, in the absence of any
forbidding principle.

To further elaborate on the second issue mentioned above, we need to
identify some observables, mostly defined in terms of asymmetries, which
subtract out the effect of CP-even operators, while keeping substantial
separation between the CP-odd ones with the standard model predictions
including their uncertainties. This is accomplished in the present
work, allowing in principle the existence of all CP-odd and-even gauge
invariant dimension-6 operators. The points we emphasize in particular
are as follows:

\begin{itemize}
\item Constraints from collider as well as EDM limits are listed in terms of
the {\em effective} $VVh$ couplings, though they can arise from all possible 
operators at the same time.
\item The contributions of the CP-even operators to the suggested asymmetries
at the LHC lie within the $1\sigma$ error-bars of the SM, while 
the CP-odd ones are shown to lead to values that steer clear of these error-bars 
over extended regions of the parameter space.
\item The prediction of observable asymmetries is based on detailed event 
generation, including showering, jet formation and detector simulation, 
thus leading to realistic estimates of what can be observed. 
\item Our `realistic' predictions of CP-violating asymmetries and suppression of
the effects of CP-even operators are presented for various luminosities, giving an idea
of the luminosity required to obtain useful conclusions.
\item Correlated interactions arising due to the gauge-invariant nature of the
parent operators are always taken into account, taking, for example, the
$\gamma Z h$ and $\gamma \gamma h$ couplings arising concomitantly 
from the dimension-6 operators (and thus competing favorably with their 
loop-induced  SM counterparts).
\end{itemize}

A resume of the CP-odd dimension-6 operators as well as their 
current limits is given in section 2. In section 3 we discuss
the observables which elicit the presence of CP-violating operators
and help distinguish them from the SM and the corresponding CP-even 
counterparts. The role of asymmetries constructed from the kinematic
distributions of such observables is also discussed.  We summarize and conclude in
section 4.


\section{CP-odd anomalous couplings and their constraints}

The  set of CP-violating gauge-Higgs
operators of dimension-6 that we have considered in our analysis  \cite{Dwivedi:2015nta} is listed below :
\begin{eqnarray}\label{eq:O-cp-odd}
 \tilde{O}_W &=& \frac{\tilde{f}_{W}}{\Lambda^2}(D_{\mu}\Phi)^\dagger\hat{\tilde{W}}^{\mu\nu}(D_{\nu}\Phi);~~ 
 \tilde{O}_B = \frac{\tilde{f}_{B}}{\Lambda^2}(D_{\mu}\Phi)^\dagger\hat{\tilde{B}}^{\mu\nu}(D_{\nu}\Phi); \nonumber \\
 \tilde{O}_{BB} &=& \frac{\tilde{f}_{BB}}{\Lambda^2}\Phi^\dagger\hat{B}^{\mu\nu}\hat{\tilde{B}}_{\mu\nu}\Phi;~~~~~~ 
 \tilde{O}_{WW} = \frac{\tilde{f}_{WW}}{\Lambda^2}\Phi^\dagger\hat{\tilde{W}}^{\mu\nu}\hat{W}_{\mu\nu}\Phi \nonumber; \\
 \tilde{O}_{BW} &=& \frac{\tilde{f}_{BW}}{\Lambda^2}\Phi^\dagger\hat{\tilde{B}}^{\mu\nu}\hat{W}_{\mu\nu}\Phi,
 \end{eqnarray}
 
 Here $\Lambda$ is the cut-off scale above which the high scale physics sets in. 
For the purpose of this study $\Lambda$ is taken to be 1 TeV.
Since for the LHC observables, $\tilde fi/{\Lambda^2}$ ($i = B, W, WW $ etc. )
is the effective parameter that gets constrained, the global fit constraints can be 
predicted for any value of $\Lambda$, with the constraints on the parameters $\tilde fi$'s
being appropriately scaled. However, for the constraints coming from EDM measurements, 
one finds that EDMs have an explicit cut-off dependence other than the one coming 
via the parameters  $\tilde fi/{\Lambda^2}$. This dependence coming from loop
contributions is logarithmic in nature \cite{Dwivedi:2015nta}. \\
However, all five of these CP-violating operators are not independent, and are related by the following constraint equations, which arise
from using the SM equations of motion: \footnote{We thank Wouter Dekens and Jordy de Vries for helpful comments on this matter.}
\begin{eqnarray}\label{eq:operator_constraints}
2\tilde{O}_B &=& \frac{\tilde{f}_{B}}{\tilde{f}_{BB}} \tilde{O}_{BB} + \frac{\tilde{f}_{B}}{\tilde{f}_{BW}} \tilde{O}_{BW},\nonumber \\
2\tilde{O}_W &=& \frac{\tilde{f}_{W}}{\tilde{f}_{WW}} \tilde{O}_{WW} + \frac{\tilde{f}_{W}}{\tilde{f}_{BW}} \tilde{O}_{BW} 
\end{eqnarray}
The above set of constraints in Eq. \ref{eq:operator_constraints} entails only three of the five operators in Eq. \ref{eq:O-cp-odd} to be
independent of each other. For the purpose of our study we take  $\tilde{O}_B$, $\tilde{O}_W$ and $\tilde{O}_{WW}$ 
to be comprising an independent set of CP-violating gauge-Higgs operators.
These operators give rise to the CP-odd $VVh$ interaction vertices of following form, 
\begin{eqnarray}
\tilde{\cal L}_{VVh} &=& \frac{g m_W}{\Lambda^2}
\tilde{C}_{VVh}~ [\epsilon_{\mu\nu\alpha\beta}k_1^\alpha k_2^\beta]~ V^{\mu}(k_{1})V^{\nu}(k_{2})h(k)\label{eq:L-vvh}, 
  \end{eqnarray}
where $V = W,Z,\gamma$. The coefficients $\tilde{C}_{VVh}$ denote the effective coupling strengths of CP-odd $VVh$ couplings
and, are listed in Table 1. The $VVh$ interaction vertices arising from corresponding CP-even operators are derived in 
Ref.~\cite{GonzalezGarcia:1999fq}.

Our aim in the present work is to include scenarios where both CP-odd and-even operators 
are present, and to discriminate between them via appropriately defined asymmetry 
parameters. We demonstrate this by turning on the CP-even operator,
 \begin{eqnarray}\label{eq:O-cp-even}
{O}_{WW} = \frac{{f}_{WW}}{\Lambda^2}\Phi^\dagger\hat{{W}}^{\mu\nu}\hat{W}_{\mu\nu}\Phi,
 \end{eqnarray}      

which contributes to all the $VVh$ vertices. The additional
interaction terms thus playing a role in our analysis have the following form \cite{GonzalezGarcia:1999fq}

\begin{eqnarray}
 {\cal L}_{VVh} &=& \frac{g m_W}{\Lambda^2} C_{VVh}~ 
  [{k_1}^\nu {k_2}^\mu -  g^{\mu\nu}(k_1.k_2)]~ V_{\mu}(k_1)V_{\nu}(k_2)h(k)\label{eq:cpeven:L-wwh}.   
\end{eqnarray}

The coupling strengths of CP-even $VVh$ couplings are given by 
 \begin{eqnarray}
  C_{WWh} = -2 f_{WW};~ C_{ZZh} = -2 c_W^2 f_{WW};~ C_{\gamma\gamma h} = -2 s_W^2 f_{WW};~ C_{\gamma Zh} = -2c_W s_W f_{WW},  
 \end{eqnarray}
which are similar to the corresponding CP-odd couplings receiving contributions from $\tilde{O}_{WW}$ operator.


\begin{table}[t]
\begin{center}
\begin{tabular}{|c|c|}
  \hline
    & \\
  {\bf Coupling} & {\bf Effective coupling strength} \\
  & \\
  \hline
    & \\
  $\tilde{C}_{WWh}$    &    $(-\tilde{f}_{W}~-~2\tilde{f}_{WW})$\\
    & \\
  \hline 
   & \\
  $\tilde{C}_{ZZh}$ & $- 1/c^2_{W} \Big(c^2_{W}\tilde{f}_W+s^2_{W}\tilde{f}_B + 
  2c^4_{W}\tilde{f}_{WW}\Big)$ \\
   & \\
  \hline
    & \\
    $\tilde{C}_{\gamma\gamma h}$ & $-2s^2_{W}\tilde{f}_{WW}$\\
      & \\
    \hline
      & \\
  $\tilde{C}_{\gamma Zh}$ & ${t_{W}}/{2}\Big(-\tilde{f}_W + \tilde{f}_B-4 c^2_{W}\tilde{f}_{WW}\Big)$\\
    & \\
      \hline 
        \end{tabular}
        \caption{ CP-odd $VVh$  coupling factors and their effective strengths in terms of operator 
        coefficients $\tilde{f}_{B}, \tilde{f}_{W}$  and $ \tilde{f}_{WW}.$ }
        \label{tab:Anomalous_vertices}
        \end{center}
 \end{table}


In \cite{Dwivedi:2015nta} we derived limits on the parameters of the CP-odd operators 
using the collider data (from LEP and LHC) and measurements of electron and neutron EDMs. These limits 
were obtained taking mainly two parameters (2P) and three parameters (3P) nonzero at 
a time. The limits on parameters were further used to constrain the effective CP-odd $VVh$ 
couplings which are listed in Table 2.   

\begin{table}[t]
 \begin{center}
\begin{tabular}{|c|c|c|c|c|}
  \hline
  {\bf Couplings}  & \multicolumn{2}{|c|}{\bf LHC data}& \multicolumn{2}{|c|}{\bf EDM}\\
  \cline{2-5}
  &{2P case} &{3P case} &{2P case} &{3P case} \\
  \hline
    & & & &\\
  $|\tilde{C}_{WWh}|$               & $0~-~60$   &  $0~-~60$    & $ 0~-~0.17 $ & $0~-~55$  \\
    & & & &\\
  \hline 
   & & & &\\
  $|\tilde{C}_{ZZh}|$               & $25~-~80$ & $25~-~80$      & $0.11~-~0.20$ & $0.15~-~33$  \\
    & & & &\\
    \hline    
   & & & &\\
  $|\tilde{C}_{\gamma\gamma h}|$    & $0~-~0.8$  & $0~-~0.8$       & $0~-~0.16$ & $0.02~-~52$ \\
    & & & &\\
  \hline
   & & & &\\
  $|\tilde{C}_{\gamma zh}|$         & $15~-~25$  & $15~-~25$      & $0.03~-~0.25$ & $0.05~-~110$ \\
    & & & &\\
  \hline 
   \end{tabular}
  \end{center}
  \caption{Limits on CP-odd coupling strengths from LHC data and EDM measurements
  for $\Lambda$=1 TeV. 2P and 3P stand for two parameter nonzero and three parameter
nonzero cases respectively.}
  \label{tab:coupling strengths}
  \end{table}
As has been already stated, these constraints can get relaxed when we
keep a linearly independent set of CP-violating gauge-Higgs operators to be non-zero at the same time
along with the other fermion-Higgs dimension-6 operators (which contribute to fermionic EDMs),
and also include at least some of their CP-conserving
counterparts. Such operators can be obtained directly by extending those listed in \cite{Grzadkowski:2010es}, Table 2.
 While the CP-violating ones contribute to EDMs, they do not affect the phenomenology of $VVh$ interactions.
 The consequent  enlargement of the parameter space, however, relaxes the otherwise stringent constraints from EDMs. 

The allowed values of the coefficients of the dimension-6 operators apparently threaten perturbative unitarity
in $V_L V_L \rightarrow V_L V_L (V = W, Z)$ at scales of the order of few TeV. 
The exact value of this scale depends on the specific choice of the operators  involved and the corresponding anomalous
couplings they generate \cite{Choudhury:2012tk, Dahiya:2013uba}. However, the effective operators are not reliable at such scales.
There one has to include the actual heavy fields of mass ${\cal O}(\Lambda)$ (for weakly coupled 
ultraviolet completion), which are integrated out to obtain the effective coupling terms. 
These additional degrees of freedom are trusted to restore unitarity when the full theory
is switched on.
In the context of the study presented here, the range of $\tilde C_{ZZh}$ is varied
upto  ${\cal O}(70)$. Even though $\tilde C_{ZZh}$  is varied upto this range, the effective coupling
parameter that will decide the perturbativity of the loop expanson is given by \cite{Christensen:2010pf},
\begin{equation}
 \tilde b = \frac{gm_{W}v}{2\Lambda^2} \tilde C_{ZZh}
\end{equation}
Here, $v = 246$ GeV is the Higgs vacuum expectation value.
Thus even for $\tilde C_{ZZh} \lesssim {\cal O}(70)$, $ \tilde b  \lesssim {\cal O}(1)$ and the 
loop expansion in terms of $\tilde b$ remains perturbative.

\section{Isolating the effect of CP-odd couplings}

 In any elementary particle scattering process, the size and shape of certain kinematic 
distributions involving leptons and jets can change non-trivially with respect to the 
purely standard model prediction in the presence of both CP-odd and-even higher dimensional 
operators.
Some of these have been studied mostly in purely phenomenological formulations, i.e., without
 working in terms of gauge invariant dimension-6 operators or without performing a realistic 
 simulation of events 
 (\cite{Djouadi:2013yb,Sun:2013yra,Anderson:2013afp,Desai:2011yj,
 Han:2000mi,Chang:1993jy,Biswal:2009ar,Biswal:2012mp,Biswal:2005fh,Godbole:2007cn}). 
For generic observables  like $p_T$ and rapidity distributions which are affected 
by both CP-even as well as CP-odd couplings, it may not be possible to separate the effect of 
the two couplings with sufficient statistical significance. In other words, for 
values of anomalous couplings, consistent with the available experimental data, 
it would require analysing a huge amount of data/events to confirm the effect of CP-odd couplings 
on top of the effect of CP-even couplings using generic observables. 
Therefore, it is more economical to work with observables
which are unambiguously sensitive to the definite CP-propertry of the couplings.

 In the present work, we are interested in CP-odd observables. 
The CP-odd observables can be categorised as $\hat{T}$-even or $\hat{T}$-odd type
\footnote{ $\hat{T}$ is the naive time reversal
  transformation that flips the sign of the particle spin and
  momenta, without interchanging the initial and final states.}. 
   Unlike the $\hat{T}$-even observables which require a non-zero absorptive/unitary phase to 
   quantify the CP-violation, the $\hat{T}$-odd observables do not require  
  any absorptive phase and can be calculated at tree-level. The triple cross products 
  constructed using the particle momenta/spin in a given process can serve as $\hat{T}$-odd 
  observables;  thus they need not be driven by any absorptive phase.
 A non-zero expectation value of such a CP-odd and $\hat{T}$-odd observable would imply CP-violation
at the Lagrangian level \cite{Valencia:1994zi}.
In practice, we work with
  asymmetries, constructed out of such observables, which are unlikely
  to be faked by the SM or other CP-even couplings via statistical
  fluctuation. In this section, we quantitatively establish
such characteristics of the chosen observables after a realistic
simulation of a suitable process at the LHC.


\subsection{Signal and backgrounds}
With the above considerations in view, we take up the analysis of the process, $p p
\rightarrow Z h \rightarrow \ell^+ \ell^- b \bar{b} $ with $\ell = e,
\mu$.  We consider the decay of Higgs to $ b \bar{b}$ final state to 
ensure sufficient event rates.
The final state is required to have two opposite sign leptons and {\it exactly two $b$-tagged jets}.
The proton-proton collisions are simulated at the
center-of-mass energy of 14 TeV. 
In presence of dimension-6 operators, this process also receives contribution from 
a tree-level diagram involving $\gamma Zh$ vertex, which is not present in the SM.
Thus there is an interference of the two $s-$channel diagrams involving the  $Z Zh$ 
and  $\gamma Zh$ vertices at the amplitude level.
This is because the same gauge-Higgs
operators that give rise to the anomalous $C_{ZZh}$ coupling also give
rise to the $C_{\gamma Z h}$ interaction [Table \ref{tab:Anomalous_vertices} ].

Throughout our analysis we consider $M_h = 126$ GeV. 
 We implement the anomalous couplings in {\tt Madgraph}
\cite{Alwall:2014hca} using the package {\tt FeynRules}
\cite{Christensen:2008py}. The cross sections are calculated using
{\tt nn23lo1} \cite{Ball:2012cx} parton distribution functions with default settings
for renormalization and factorization scales.
     While the parton level events are generated in
     {\tt Madgraph}, the Pythia \cite{Sjostrand:2006za} switch
     is turned on to incorporate showering, hadronisation and initial
     and final state radiation (ISR/FSR) effects. The HEP file
     obtained from this interfacing is passed to {\tt Delphes-3.1.2}
     \cite{deFavereau:2013fsa} which gives a ROOT \cite{root_package}
     file as output. Jets are formed using the anti-kt algorithm in {\tt Delphes}
     with a cone size of 0.6.
     A $p_T$ dependent tagging efficiency is employed for $b$-jets, namely, 
     0.7-0.5 for $p_T$ in the range 20-500 GeV. 
We also use a mis-tagging efficiency of 0.001 for  light jets and 
a $p_T$ dependent mis-tagging efficiency for $c$-induced jets, which varies from
0.14-0.08 for 20 $< p_T <$  500 GeV. 
     The ROOT file thus obtained is fed to the analysis code to generate the
     distributions.  \\

We start with following basic cuts on the transverse momentum, rapidity and
separation of leptons ($\ell$) and jets ($J$), where $J~=~ j, b$; $j$ being
the light quark jets and $b$ corresponds to the $b$-induced jets.\\
\begin{enumerate}
 \item $p_{T_{J, \ell}}>$  20 GeV
  and $|\eta_{J,\ell}|<$ 2.5 :  We demand that our final state leptons and jets should
lie in the central rapidity region each with a minimum transverse momentum of 20
GeV.
 \item $\Delta R_{JJ,J\ell}>$  0.4 : We demand that both the jets should be well
 separated from each other as well as from leptons.
 \item $\Delta R_{\ell \ell}>$ 0.2 : There will be sufficiently large opening angle between the two leptons.
\end{enumerate}

The SM cross section  for the signal after the generation level cuts is 12.4 fb.

The major background to the signal comes from the following SM processes:\\
\begin{enumerate}
 \item $p p \rightarrow ~  b \bar{b}~ \ell^+ \ell^- ~({\rm QCD})$
 \item $p p \rightarrow t \bar{t} \rightarrow  b \bar{b}~ \ell^+ \ell^- ~\nu \bar{\nu}$
 \item $p p \rightarrow  j j ~\ell^+ \ell^-~ ({\rm QCD})$
 \item $ p p \rightarrow Z~\ell^+ \ell^- \rightarrow  b \bar{b}~ \ell^+ \ell^-$ (EW)
\end{enumerate}

To suppress the above sources of the background, we employ the following additional
selection cuts :\\
\begin{enumerate}
 \item $p_{T_{J}}>$  50 GeV
 \item 105$<M_{JJ}<$ 130 GeV
 \item $|M_{ll} - M_Z| <$ 15 GeV
 \item The $p_T$ of the reconstructed $Z$ boson is taken to be $>$ 150 GeV
\end{enumerate}
The asymmetrical cut on the invariant mass  of the two $b$-tagged jets ($M_{JJ}$) in the final state
around the peak ( approximately 116 GeV) is helpful in suppressing backgrounds from
$p p \rightarrow  b \bar{b}~ \ell^+ \ell^-$ and 
$ p p \rightarrow Z~\ell^+ \ell^- \rightarrow  b \bar{b}~ \ell^+ \ell^-$ processes.
We have taken such a window for the invariant mass of the two $b$-tagged jets because 
owing to the effects of hadronisation and showering, the peak of the invariant mass
distribution shifts from $M_h$ to about 116 GeV. Thus we gain in terms of signal events by choosing
such an invariant mass cut.
Further, requiring exactly two $b$-tagged jets in the final state significantly reduces
the background coming from $p p \rightarrow   j j ~\ell^+ \ell^-$.

 In order to suppress the $t \bar{t}$ backgrounds, use has been made of the fact that our signal
events are hermetic in the ideal situation. In contrast, similar final states from the
$t \bar{t}$ background will be necessarily associated with missing-$E_T$. Thus we propose
to distinguish the signal by demanding  $ \cancel{E}_{T}<$  20 GeV \cite{Christensen:2010pf}.
Moreover, no missing-$p_T$ pointing centrally is ensured by choosing  $|\cancel{\eta}|>$  2.5,
where  $\cancel{\eta}$ stands for the polar direction of the transverse momentum imbalance.

\subsection{Observables}
Next, we discuss two CP-odd, $\hat{T}$-odd quantities used in the rest of the analysis 
 to isolate the effect of CP-odd couplings from the CP-even ones. These observables
are constructed using the 3-momentum information of the final state particles. The first one
, also discussed in Refs.~\cite{Han:2009ra,Christensen:2010pf} is given by,
\begin{equation}
{\cal O}_1 = \frac{(\vec{p}_+ \times\vec{p}_-)\ldotp
  \hat{z}}{|\vec{p}_+ \times \vec{p}_-|}~ {\rm sign}[(\vec{p}_+ ~-~
  \vec{p}_-)\ldotp \hat{z}]
\end{equation}
 Here $\vec{p}_+$ and $\vec{p}_-$ are the 3-momenta of $\ell^+$ and
 $\ell^-$ respectively, and $\hat{z}$ is the unit vector along the
 incoming quark(or anti-quark) direction or the collision axis.  The
 factor ${\rm sign}[(\vec{p}_+ ~-~ \vec{p}_-)\ldotp \hat{z}] $ is the
 sign of the difference of the momentum projections of the outgoing
 leptons along the $\hat{z}$ axis.  Since the dot product with the
 $\hat z$ unit vector occurs twice in the definition of the
 observable, the observable is rendered independent of the choice of
 the direction of the incoming quark momentum.  This is important in
 the context of LHC, as there is ambiguity associated with the
 momentum direction of the initial quark with respect to which proton
 it is from. The definition of ${\cal O}_1$ removes this ambiguity and
 makes it uniquely measurable.

 The second observable that we consider uses  $\vec{p}_{j_1}$ and $\vec{p}_{j_2}$, the three-momenta of the two
 $b$-tagged jets in addition to the oppositely signed leptons
 in the final state and is given as:\\
 \begin{equation}
 {\cal O}_2 = \frac{((\vec{p}_{j_1}+ \vec{p}_{j_2}) \times (\vec{p}_+ ~-~ \vec{p}_-))\ldotp
   \hat{z}}{|(\vec{p}_{j_1}+ \vec{p}_{j_2}) \times (\vec{p}_+ ~-~ \vec{p}_-)|}~ {\rm
   sign}[(\vec{p}_+ ~-~ \vec{p}_-)\ldotp \hat{z}],
 \end{equation}

 These observables are CP-odd and $\hat{T}$-odd, and thus generated
 from the dispersive part of the amplitude. It should also be noted
 that, since the in-state partons from the two colliding proton beams
 do not always carry equal momenta, the lab frame and the center of
 mass (cm) frame for the colliding partons are in general
 different. Thus, except for the cm frame {\it i.e.} when the two initial
 partons come with the same $x$-value, where
 $\vec{p}_{j_1}+ \vec{p}_{j_2}~+~\vec{p}_+ ~+~ \vec{p}_- ~=~ 0$, the observables ${\cal O}_1$ and
 ${\cal O}_2$ are distinct. Also, due to the showering effects, an exact 
 momentum balance between the final state $b$-jets and leptons does not
 hold true. The difference in the observables is evident from their distributions. 
 By definition, each of them ranges from -1.0 to +1.0.

 The distinguishability of the CP-odd operators is quantified more effectively
in terms of the asymmetries $A_i$ defined as:
 \begin{equation}
 A_i = \frac{\sigma({\cal O}_i > 0) ~-~ \sigma({\cal O}_i < 0) }{\sigma({\cal O}_i > 0) ~+~ \sigma({\cal O}_i < 0) } \label{eqn:Asymmetry_def}
\end{equation}
%

\begin{figure}[ht!]

\begin{subfigure}[b]{0.33\textwidth}
 \includegraphics[width = \textwidth]{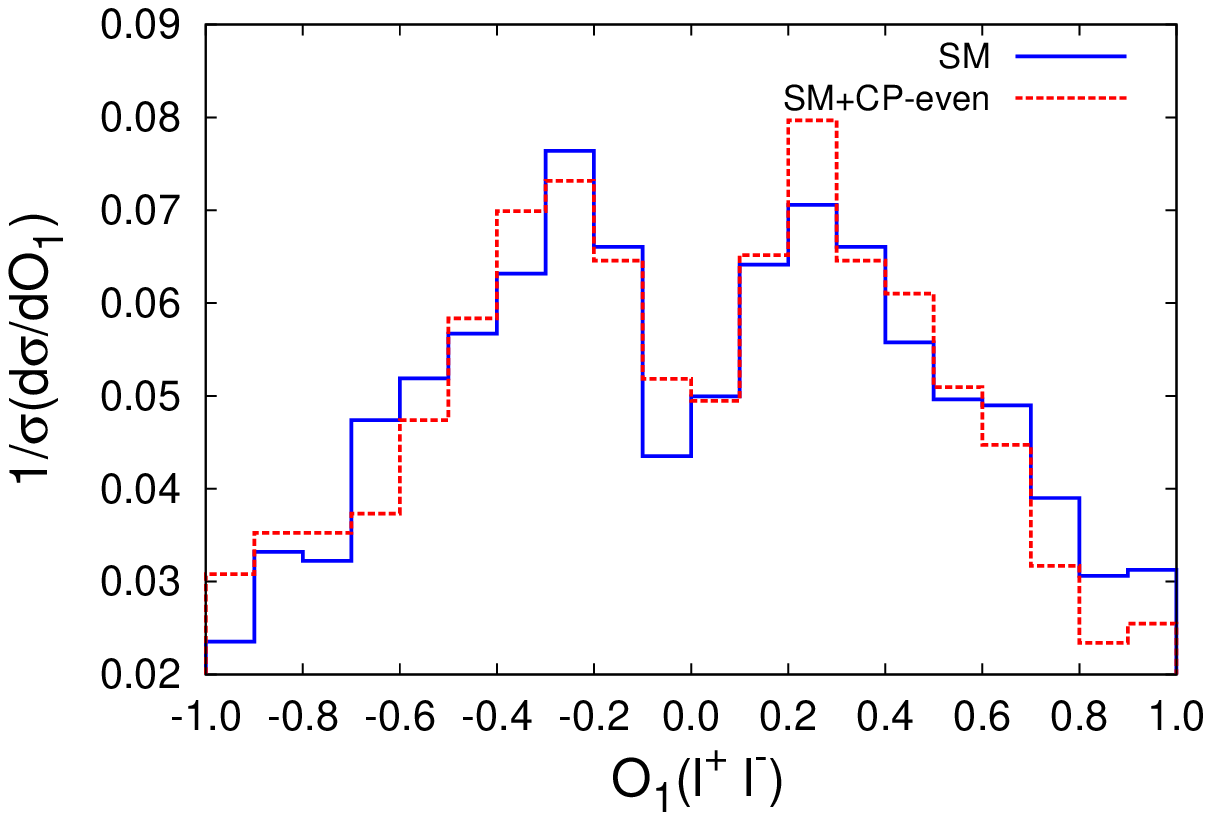}
    \caption{}\label{fig:O_1:cpeven}
    
    \end{subfigure}
   \begin{subfigure}[b]{0.33\textwidth}
  \includegraphics[width = \textwidth]{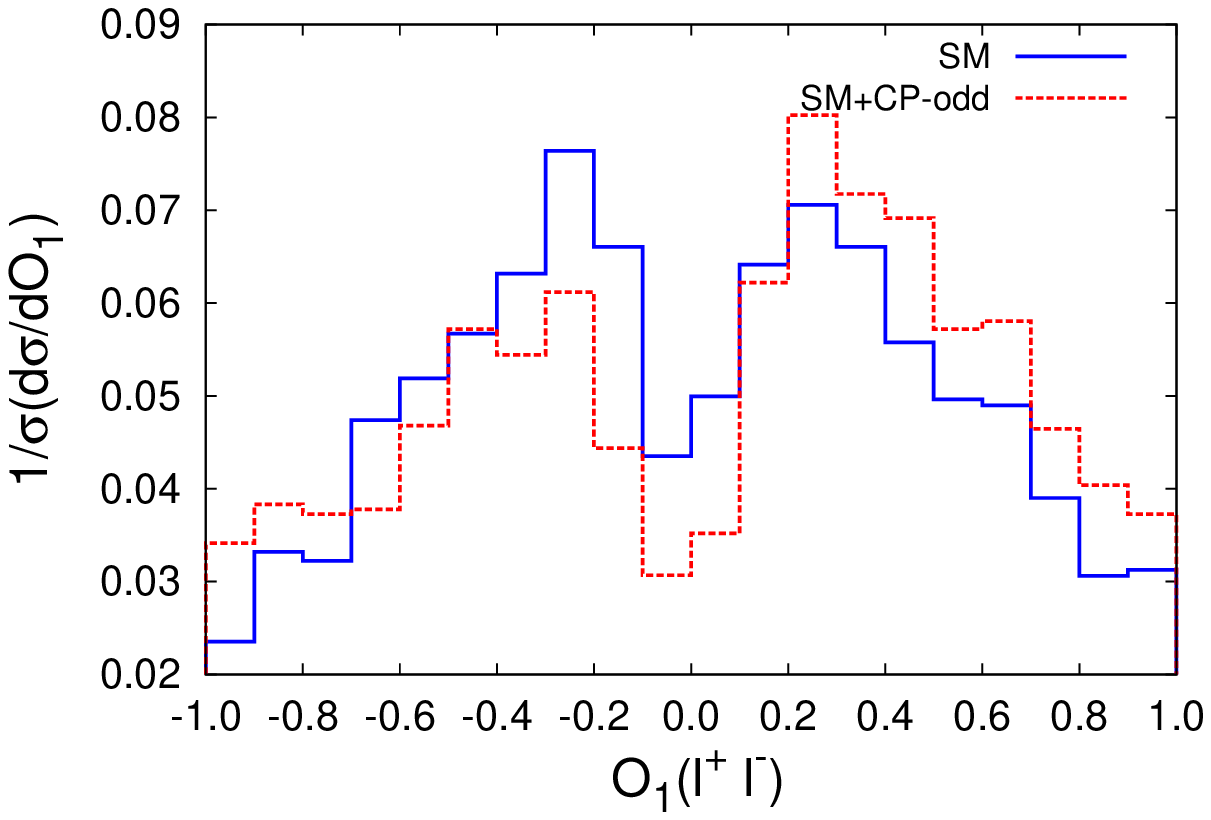}
     \caption{}\label{fig:O_1:cpodd}
   
   \end{subfigure}
    \begin{subfigure}[b]{0.33\textwidth}
  \includegraphics[width = \textwidth]{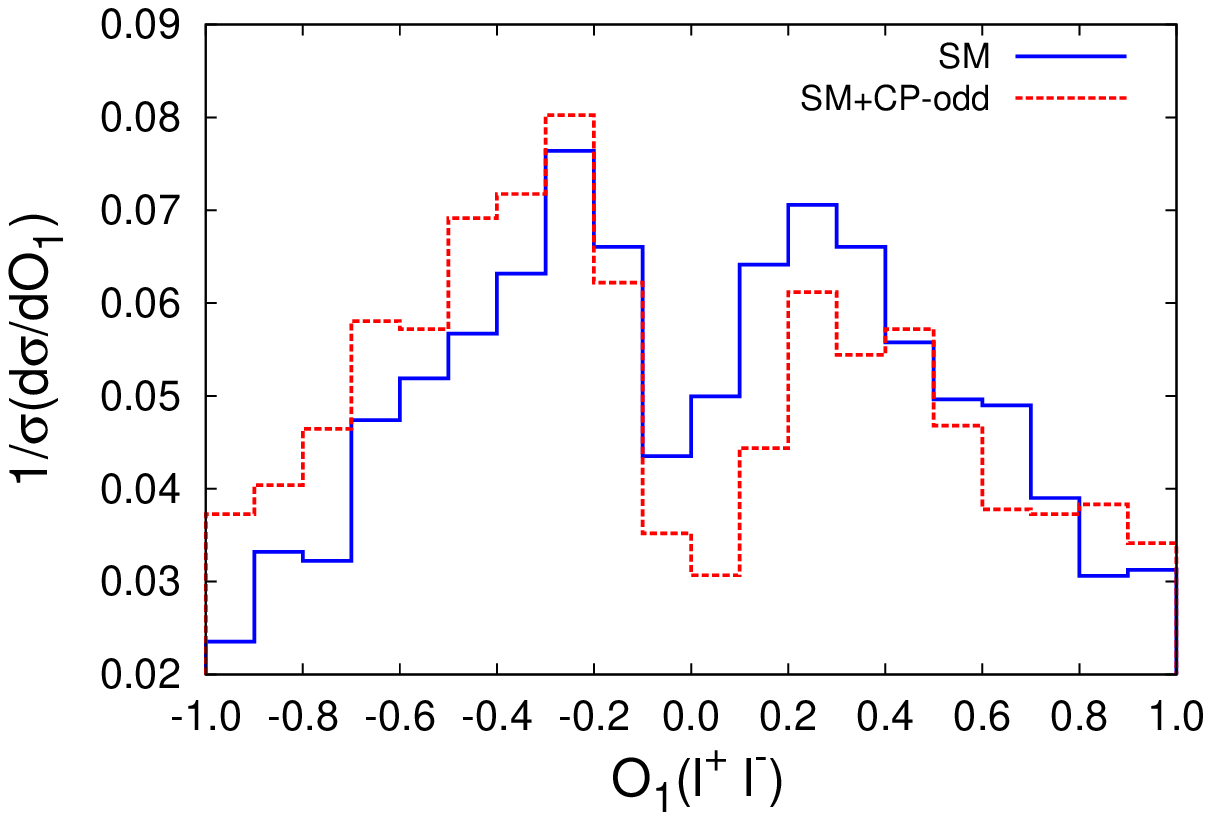}
     \caption{}\label{fig:O_1:cpodd_sign_rev}
   
   \end{subfigure}
   
   \caption{Differential cross-section distributions for ${\cal O}_1$
     at $\sqrt{s} = 14$ TeV. (a) SM vs SM + CP-even
     case with ${C}_{ZZh} = 4.62$ and ${C}_{\gamma Z h} = 2.53$. (b)
      SM vs SM + CP-odd case with $\tilde{C}_{ZZh} =
     40.0$ and $\tilde{C}_{\gamma Z h} = 0.0$. (c) We 
     reverse the sign of the CP-odd coupling with $\tilde{C}_{ZZh} =
     -40.0$ and $\tilde{C}_{\gamma Z h} = 0.0$.}
  \label{fig:O_1}
 \end{figure}

\begin{figure}[h!]

\begin{subfigure}[b]{0.33\textwidth}
 \includegraphics[width = \textwidth]{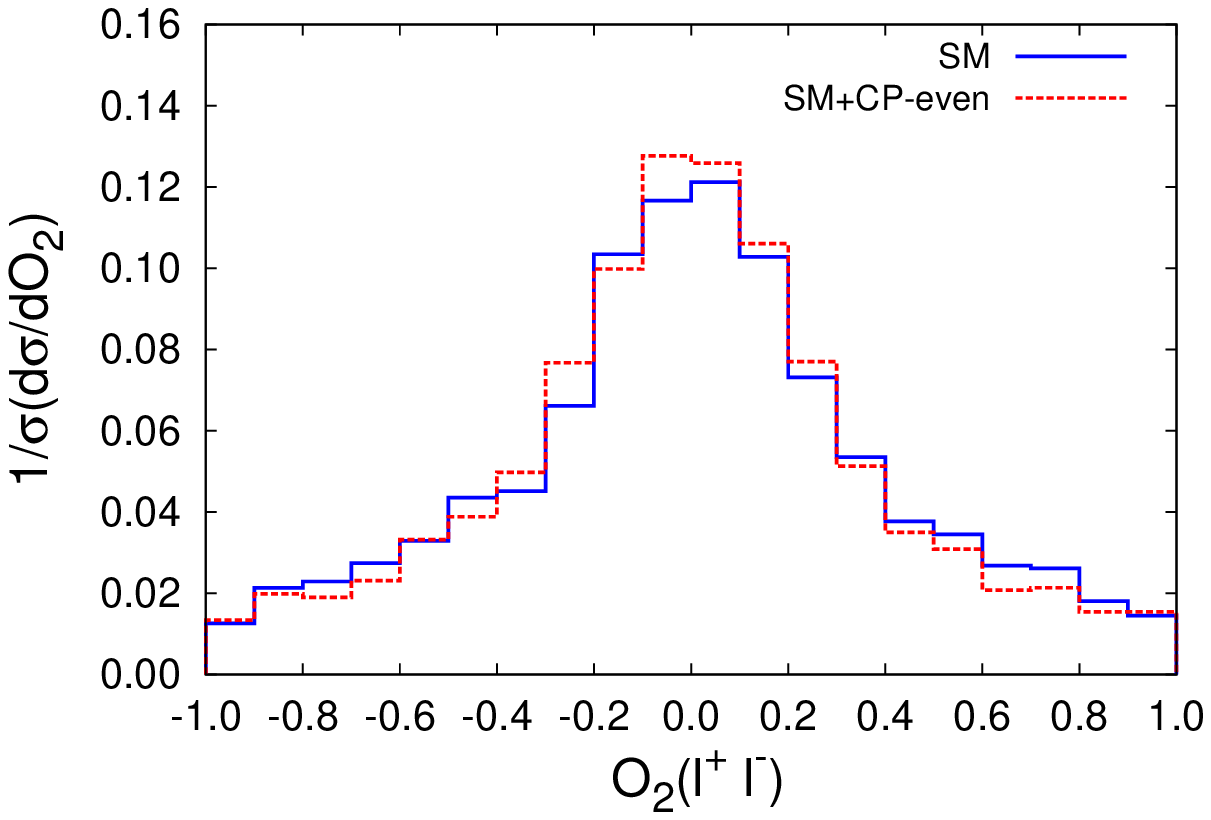}
    \caption{}\label{fig:O_2:cpeven}
    
    \end{subfigure} 
   \begin{subfigure}[b]{0.33\textwidth}
  \includegraphics[width = \textwidth]{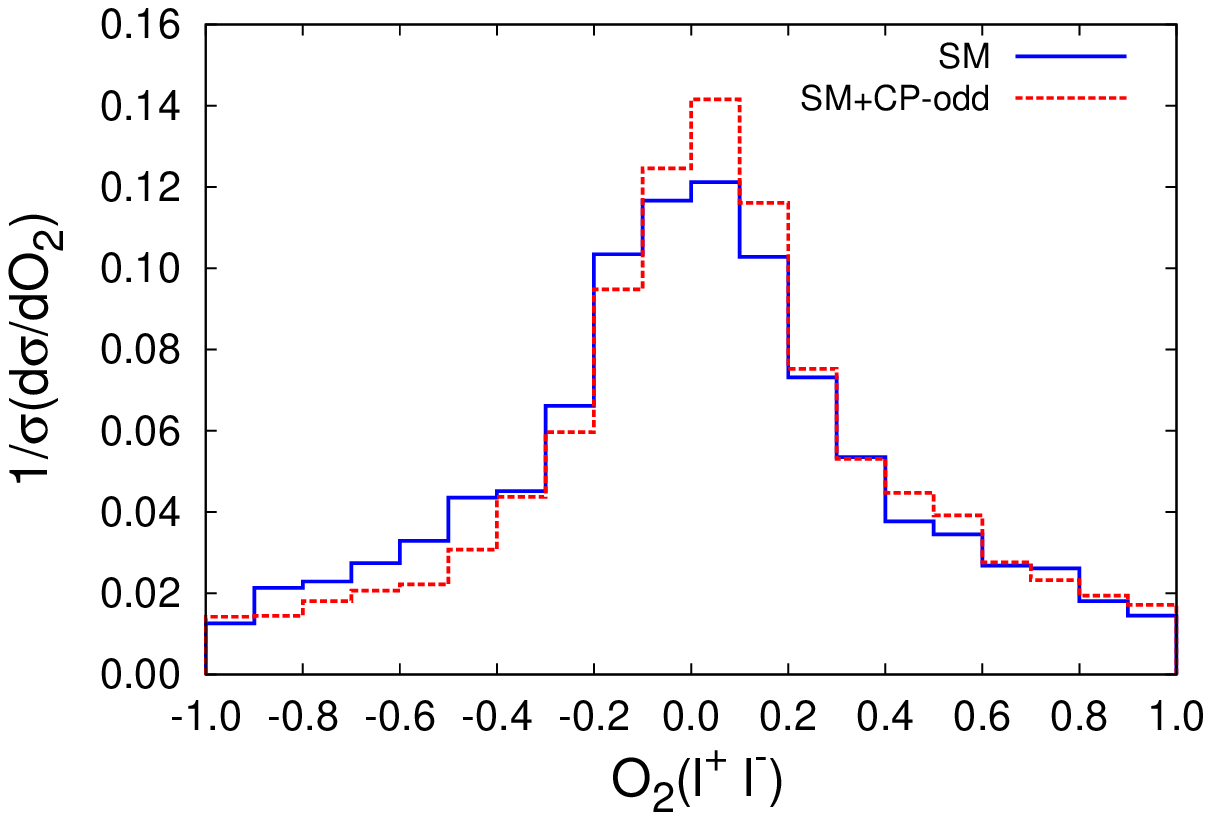}
     \caption{}\label{fig:O_2:cpodd}
   
   \end{subfigure}
    \begin{subfigure}[b]{0.33\textwidth}
  \includegraphics[width = \textwidth]{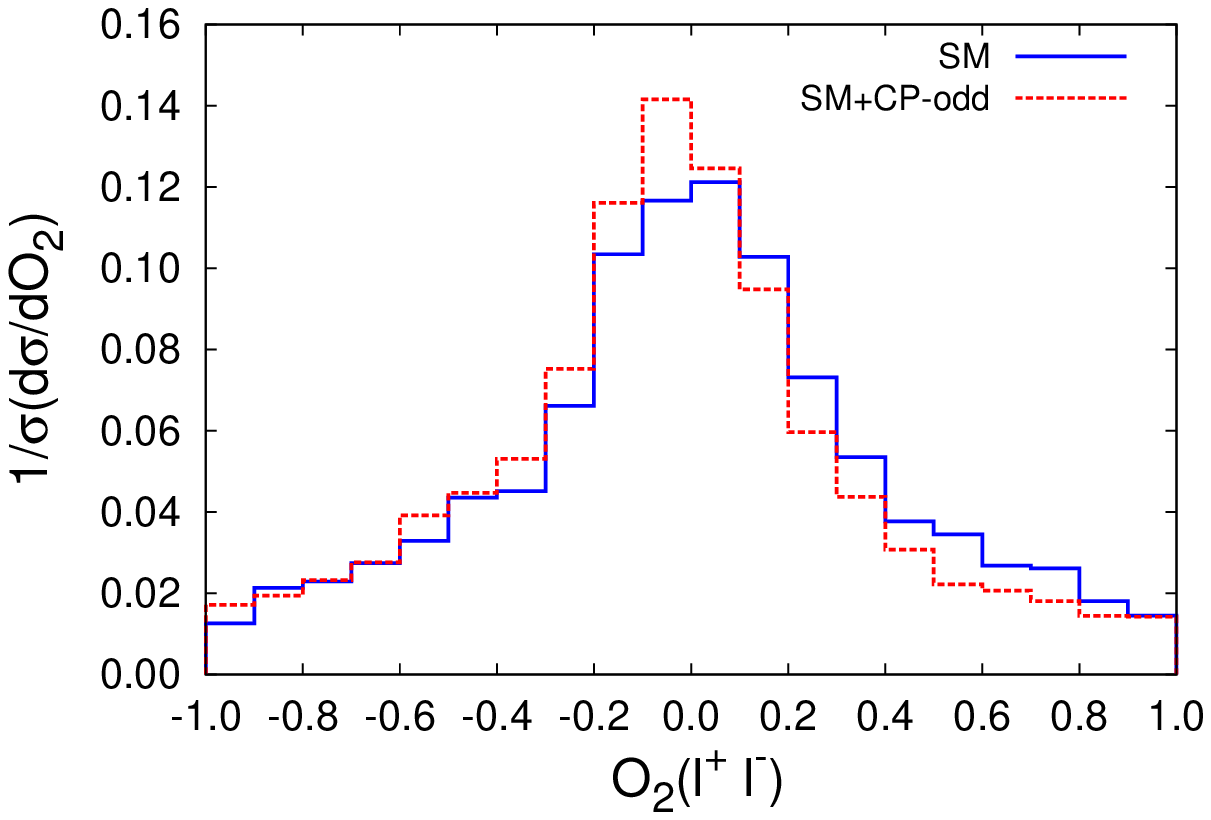}
     \caption{}\label{fig:O_2:cpodd_sign_rev}
   
   \end{subfigure}
   
   \caption{Differential cross-section distributions for ${\cal O}_2$
     at $\sqrt{s} = 14$ TeV. (a) SM vs SM + CP-even
     case with ${C}_{ZZh} = 4.62$ and ${C}_{\gamma Z h} = 2.53$. (b)
      SM vs SM + CP-odd case with $\tilde{C}_{ZZh} =
     40.0$ and $\tilde{C}_{\gamma Z h} = 0.0$. (c) We 
     reverse the sign of the CP-odd coupling with $\tilde{C}_{ZZh} =
     -40.0$ and $\tilde{C}_{\gamma Z h} = 0.0$. }
  \label{fig:O_2}
  
\end{figure}

We show in Figs. \ref{fig:O_1} and \ref{fig:O_2} histograms of the
differential cross section distributions corresponding to observables
${\cal O}_1$ and ${\cal O}_2$ respectively. In both the figures, to
tune in the CP-even Lagrangian, we have taken the benchmark value
$f_{WW} = -3.0$  \cite{Banerjee:2015bla} which corresponds to $C_{ZZh} = 4.62 $ and $C_{\gamma Zh}$ = 2.53.
As for the CP-odd terms, we take the benchmark
point as $\tilde f_B = \tilde f_W = -30.77$ such that $\tilde{C}_{ZZh} = 40.0$
and $\tilde{C}_{\gamma Z h} = 0.0$ [Table \ref{tab:Anomalous_vertices} ]. 
 Both of these benchmarks are consistent with existing constraints 
 \cite{Dwivedi:2015nta,Banerjee:2015bla}.
The histograms in the two figures are indicative of the induced asymmetry
in the distributions for both ${\cal O}_1$ and ${\cal O}_2$ when the 
CP-violating coupling is turned on. It is important to note that 
unlike total rates and other CP-even observables which depend on CP-odd couplings 
quadratically, a comparison of distributions from 
second  and third panel confirms that these observables have linear dependence on $\tilde{C}_{ZZh}$ coupling
in the numerator of the expression for the asymmetry [Eq. \ref{eqn:Asymmetry_def} ].
The qualitative features of the histograms 
remain similar when the $\tilde{C}_{\gamma Z h}$ coupling is also taken to be non-zero
and interference effects of both $\tilde{C}_{Z Z h}$ and $\tilde{C}_{\gamma Z h}$
couplings are considered.

\subsection{Asymmetries vs couplings} 
To assess the possibility of seeing the signatures of CP-violating 
physics in isolation, we need to calculate 
the statistical significance at which one can establish the existence of non-vanishing
asymmetry over and above the CP-conserving effects coming from the SM
background. If $N_S$ is the number of signal events and $N_B$ the number of
background events, then the observed asymmetry $A^{\rm obs}_i$, corresponding
to observable ${\cal O}_i$ at the detector is related to the theoretical asymmetry
in the signal distribution {\it i.e.} $A_i$, given in Eq. (\ref{eqn:Asymmetry_def})
by the following relation:\\
\begin{equation}
A^{\rm obs}_i=A_i\Big(\frac{N_S}{N_S + N_B}\Big) 
\end{equation}

The statistical error in the observed asymmetry is given by,
\begin{equation}
 \Delta A^{\rm obs}_i = \frac{1}{\sqrt{N_S + N_B}} = \frac{1}{\sqrt{N}},
\end{equation}

where $N = N_S + N_B$ is the total number of signal and background events.
Using the above information, the significance $S$, associated with an
observed asymmetry is given by \cite{Godbole:2007cn},
\begin{equation}
 S = A^{\rm obs}_i \sqrt{N} = A_i \frac{N_S}{\sqrt{N}}
\end{equation}

\begin{figure}[!tbp]

\begin{subfigure}[b]{0.33\linewidth}
 \includegraphics[width = \linewidth]{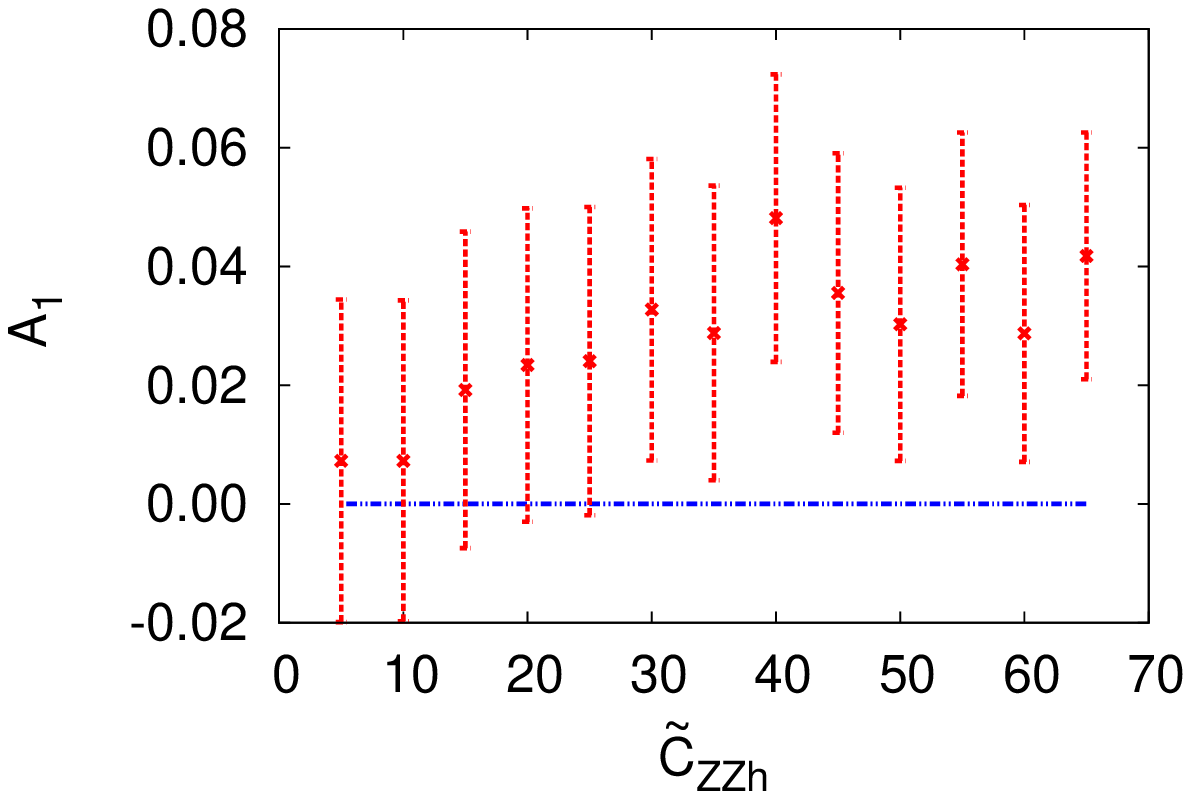}
    \caption{$\mathcal{L} = 1~ {\rm ab}^{-1}$  }\label{fig:O_1_L_1000:no_gamma_zh}
    
    \end{subfigure}
  \begin{subfigure}[b]{0.33\linewidth}
  \includegraphics[width = \linewidth]{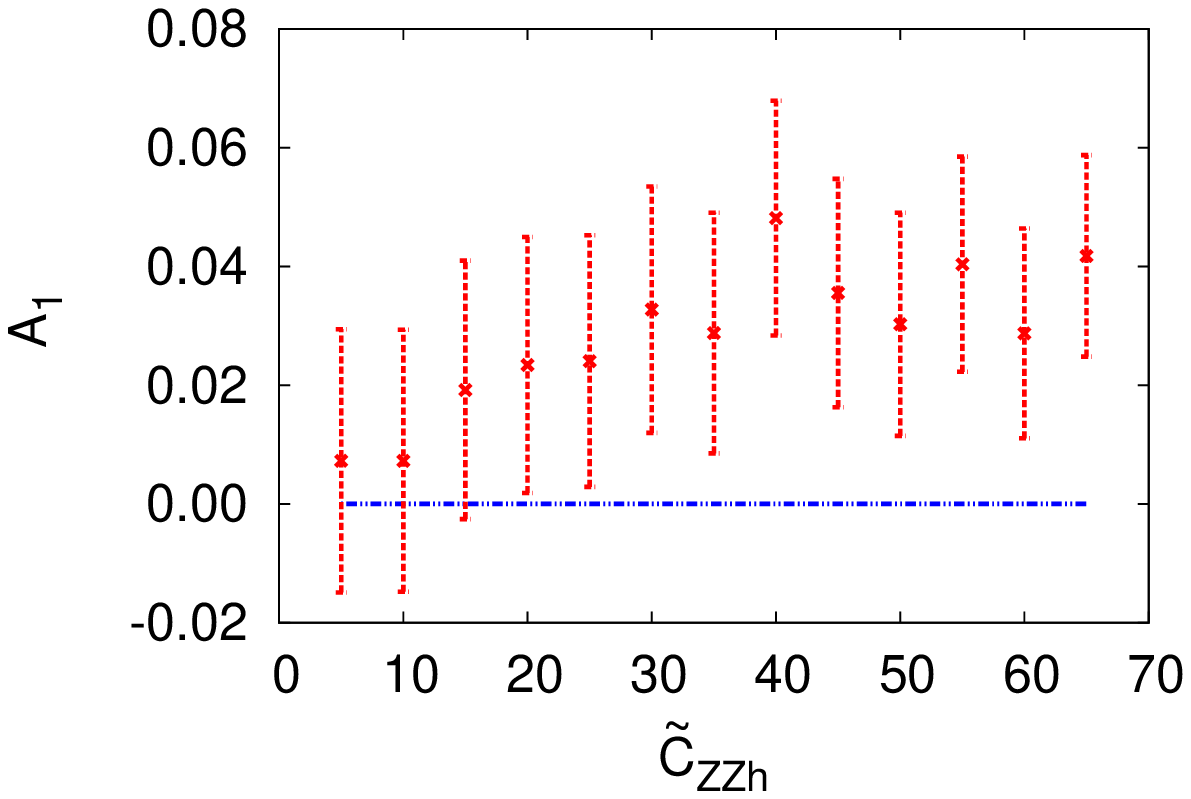}
     \caption{$\mathcal{L}= 1.5~{\rm ab}^{-1}$ }\label{fig:O_1_L_1500:no_gamma_zh}
   
   \end{subfigure}
    \begin{subfigure}[b]{0.33\linewidth}
  \includegraphics[width = \linewidth]{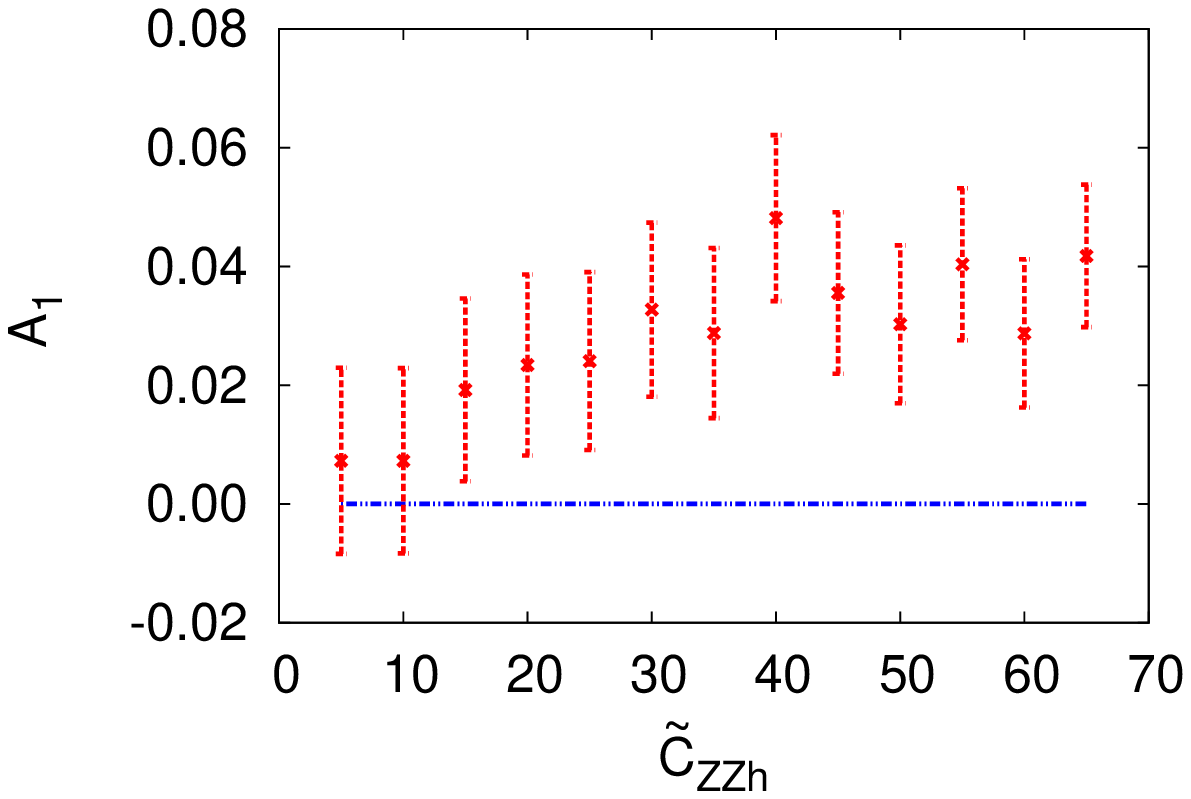}
     \caption{$\mathcal{L}= 3~{\rm ab}^{-1}$ }\label{fig:O_1_L_3000:no_gamma_zh}
   \end{subfigure}
   
    \caption{The asymmetry vs. coupling plot for the variable ${\cal O}_1$
    at Luminosities 1 ${\rm ab}^{-1}$, 1.5 ${\rm ab}^{-1}$ and 3 ${\rm ab}^{-1}$, 
   plotted for varying $\tilde{C}_{ZZh}$. The benchmark points are chosen such that
   $\tilde{C}_{\gamma Z h} = 0.0$. Statistical uncertainties are shown as error bars for 
   different benchmark points. The zero line is shown to illustrate the offset
   of the non-vanishing asymmetry against the SM and CP-even case, for which a CP-odd
   observable has no asymmetry.}
  \label{fig:O_1_asymmetry_no_gamma_zh}
 \end{figure}

\begin{figure}[!tbp]

\begin{subfigure}[b]{0.33\textwidth}
 \includegraphics[width = \textwidth]{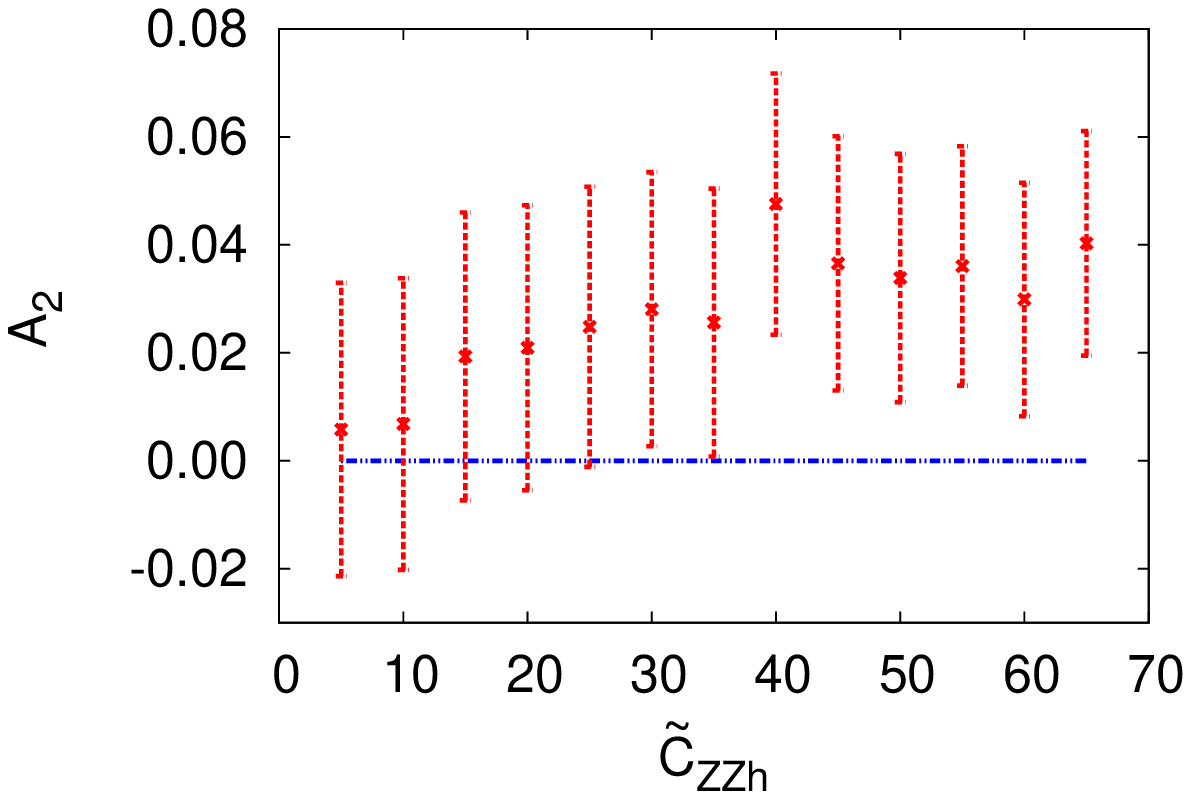}
    \caption{$\mathcal{L}= 1~{\rm ab}^{-1}$  }\label{fig:O_2_L_1000:no_gamma_zh}
    
    \end{subfigure}
  \begin{subfigure}[b]{0.33\textwidth}
  \includegraphics[width = \textwidth]{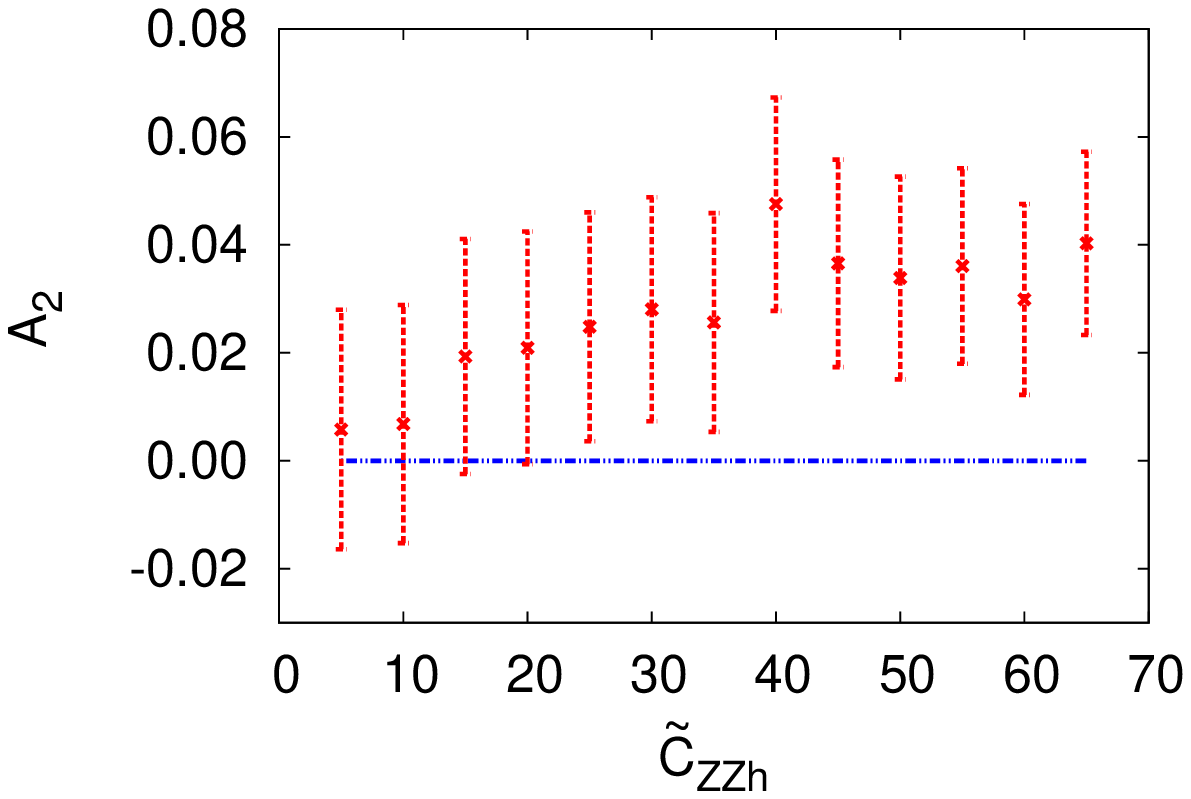}
     \caption{$\mathcal{L}= 1.5~ {\rm ab}^{-1}$ }\label{fig:O_2_L_1500:no_gamma_zh}
   
   \end{subfigure}
    \begin{subfigure}[b]{0.33\textwidth}
  \includegraphics[width = \textwidth]{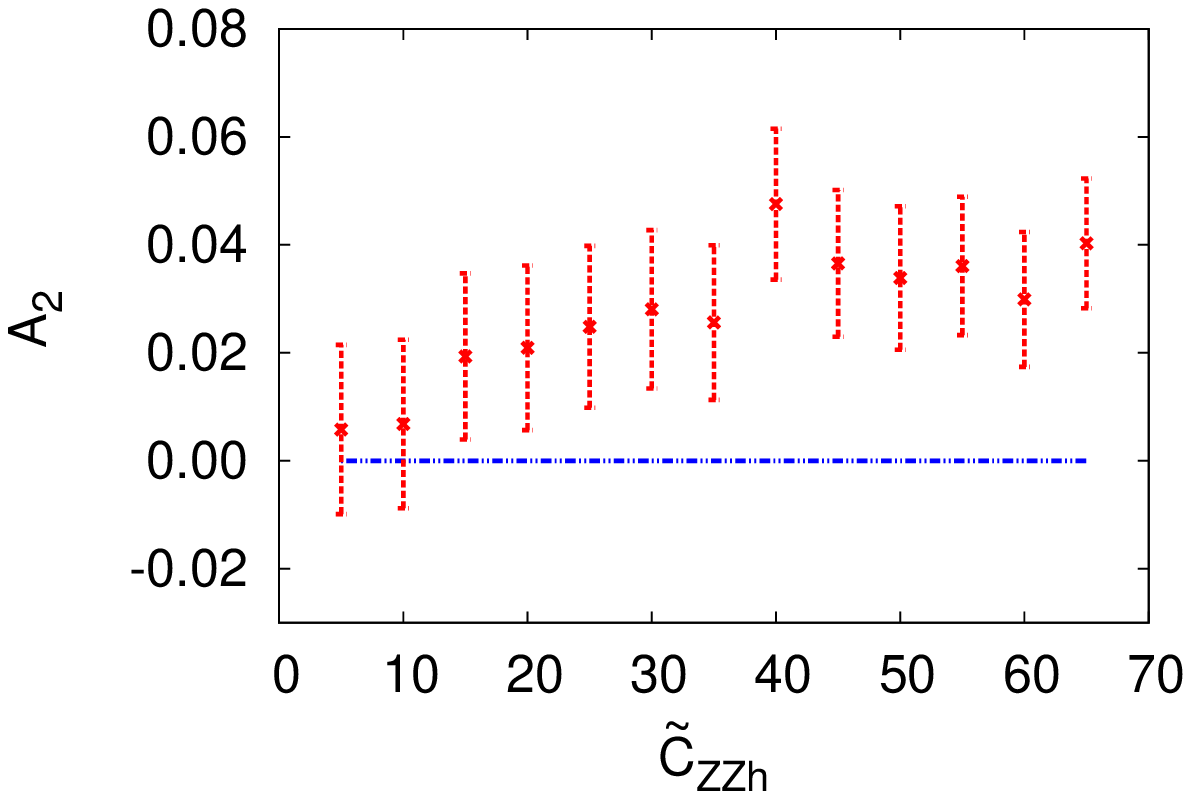}
     \caption{$\mathcal{L}= 3~{\rm ab}^{-1}$ }\label{fig:O_2_L_3000:no_gamma_zh}
   \end{subfigure}
   
    \caption{The asymmetry vs. coupling plot for the variable ${\cal O}_2$
    at Luminosities 1 ${\rm ab}^{-1}$, 1.5 ${\rm ab}^{-1}$ and 3 ${\rm ab}^{-1}$, 
   plotted for  varying $\tilde{C}_{ZZh}$. The benchmark points are chosen
   such that $\tilde{C}_{\gamma Z h} = 0.0$.   }
  \label{fig:O_2_asymmetry_no_gamma_zh}
 \end{figure}

\begin{figure}[!tbp]

\begin{subfigure}[b]{0.33\textwidth}
 \includegraphics[width = \textwidth]{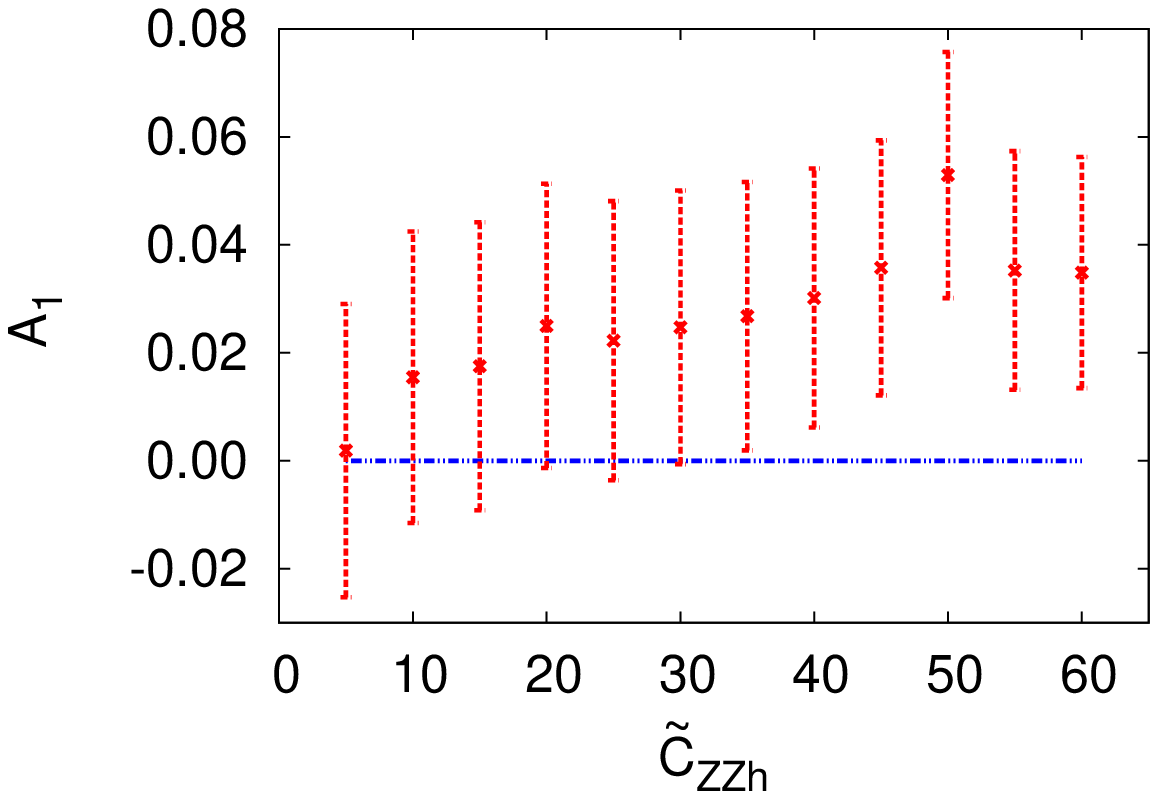}
    \caption{$\mathcal{L}= 1~{\rm ab}^{-1}$  }\label{fig:O_1_L_1000:with_gamma_zh}
    
    \end{subfigure}
  \begin{subfigure}[b]{0.33\textwidth}
  \includegraphics[width = \textwidth]{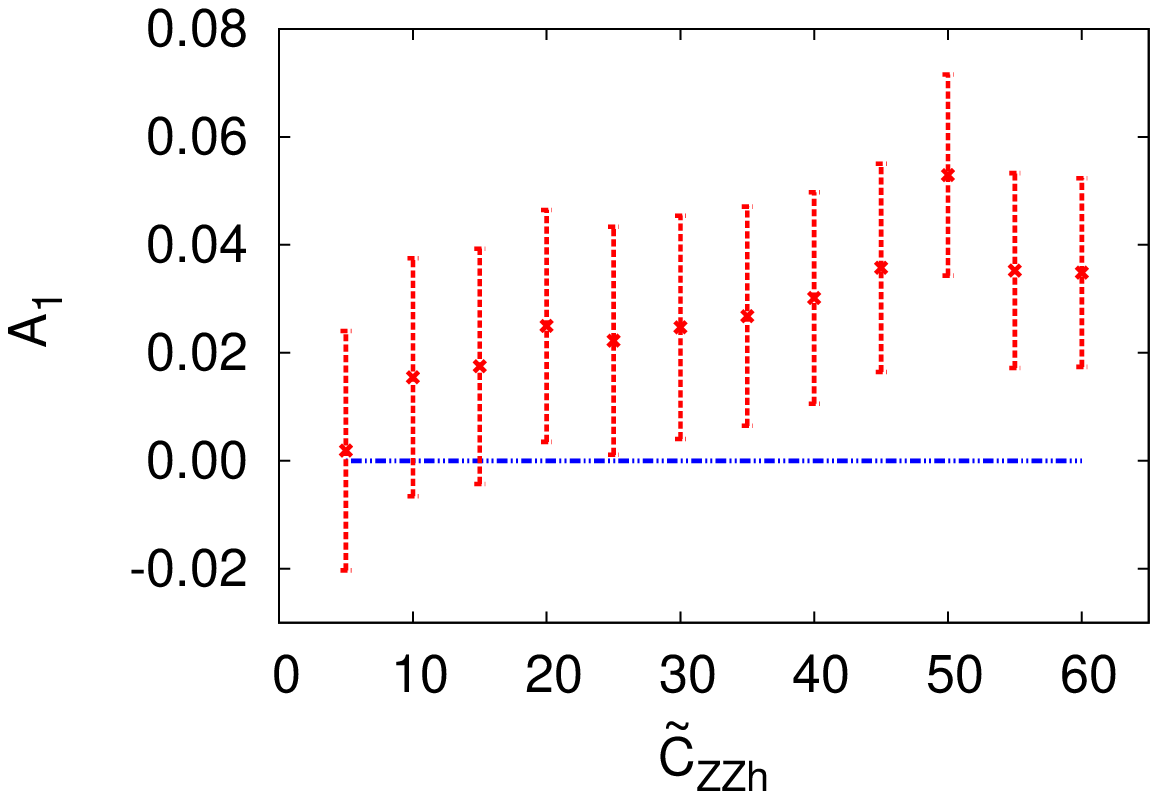}
     \caption{$\mathcal{L}= 1.5~{\rm ab}^{-1}$ }\label{fig:O_1_L_1500:with_gamma_zh}
   
   \end{subfigure}
    \begin{subfigure}[b]{0.33\textwidth}
  \includegraphics[width = \textwidth]{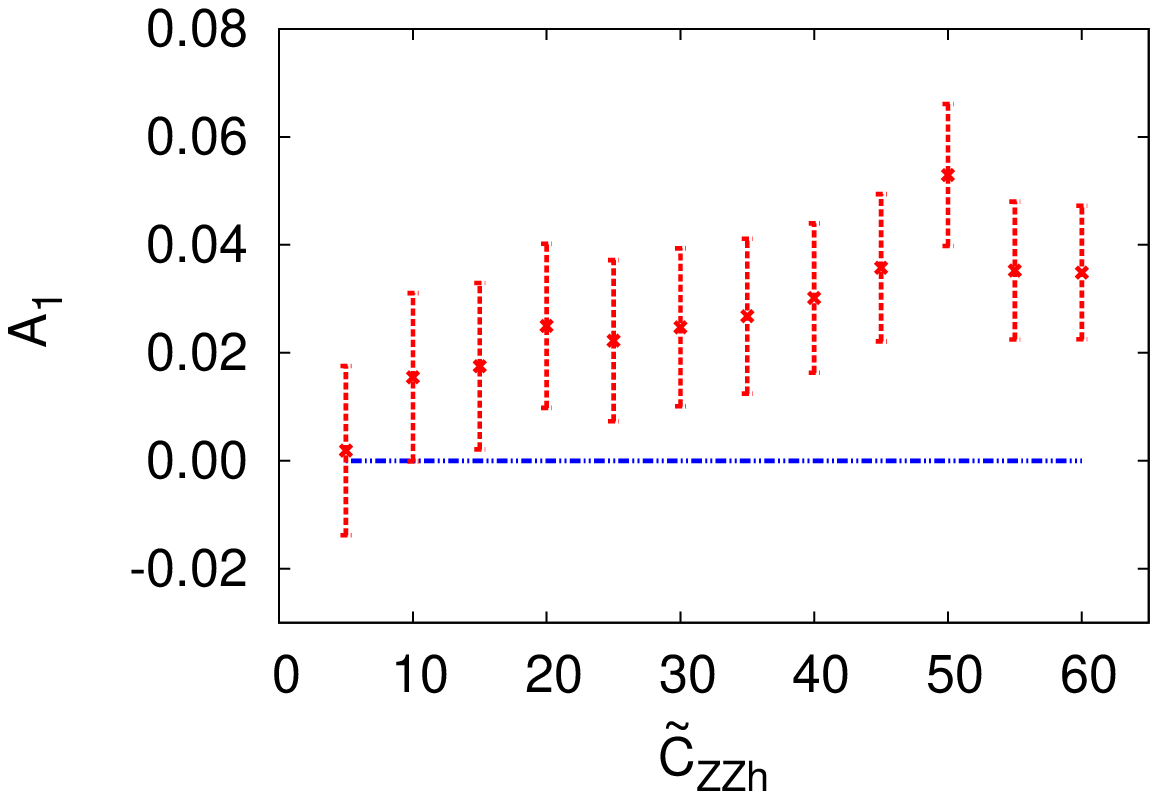}
     \caption{$\mathcal{L}= 3~{\rm ab}^{-1}$ }\label{fig:O_1_L_3000:with_gamma_zh}
   \end{subfigure}
   
    \caption{The asymmetry vs. coupling plot for the variable ${\cal O}_1$
    at Luminosities 1 ${\rm ab}^{-1}$, 1.5 ${\rm ab}^{-1}$ and 3 ${\rm ab}^{-1}$, 
   plotted for varying $\tilde{C}_{ZZh}$. The benchmark points are chosen
   such that $\tilde{C}_{\gamma Z h} = 1.5$.}
  \label{fig:O_1_asymmetry_with_gamma_zh}
 \end{figure}

\begin{figure}[!tbp]

\begin{subfigure}[b]{0.33\textwidth}
 \includegraphics[width = \textwidth]{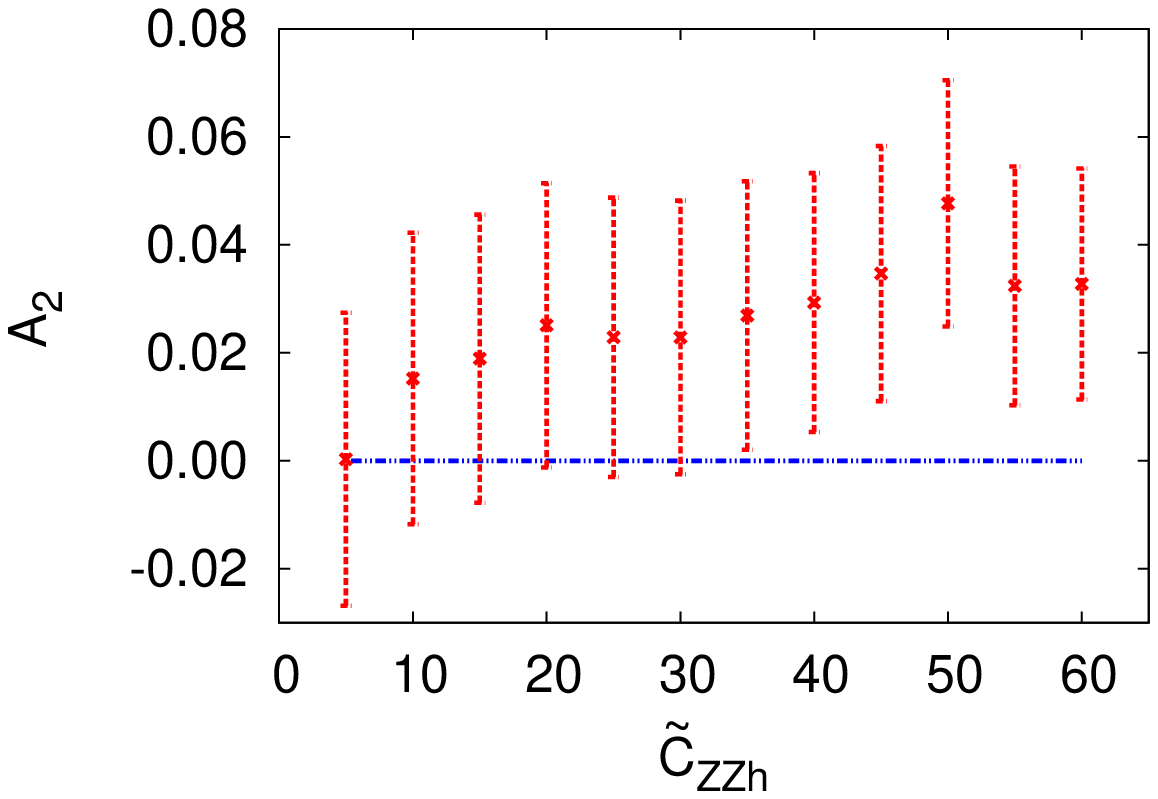}
    \caption{$\mathcal{L}= 1~{\rm ab}^{-1}$  }\label{fig:O_2_L_1000:with_gamma_zh}
    
    \end{subfigure}
  \begin{subfigure}[b]{0.33\textwidth}
  \includegraphics[width = \textwidth]{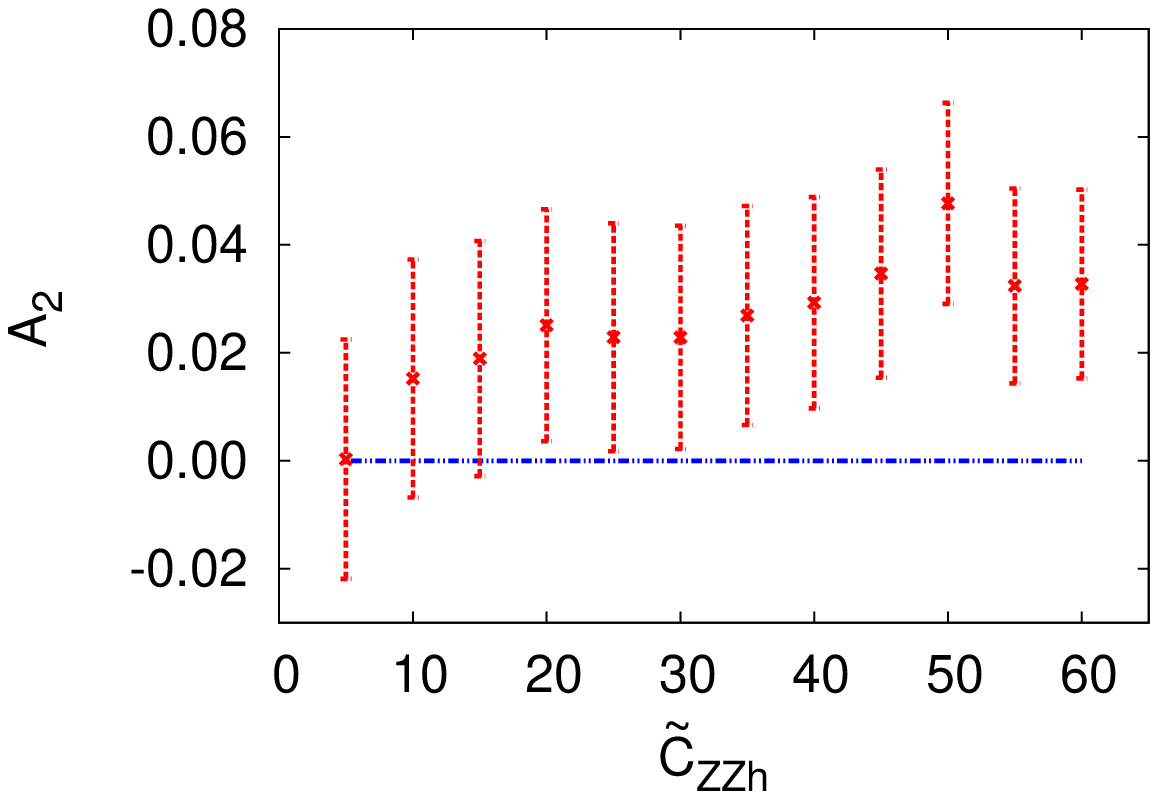}
     \caption{$\mathcal{L}= 1.5~{\rm ab}^{-1}$ }\label{fig:O_2_L_1500:with_gamma_zh}
   
   \end{subfigure}
    \begin{subfigure}[b]{0.33\textwidth}
  \includegraphics[width = \textwidth]{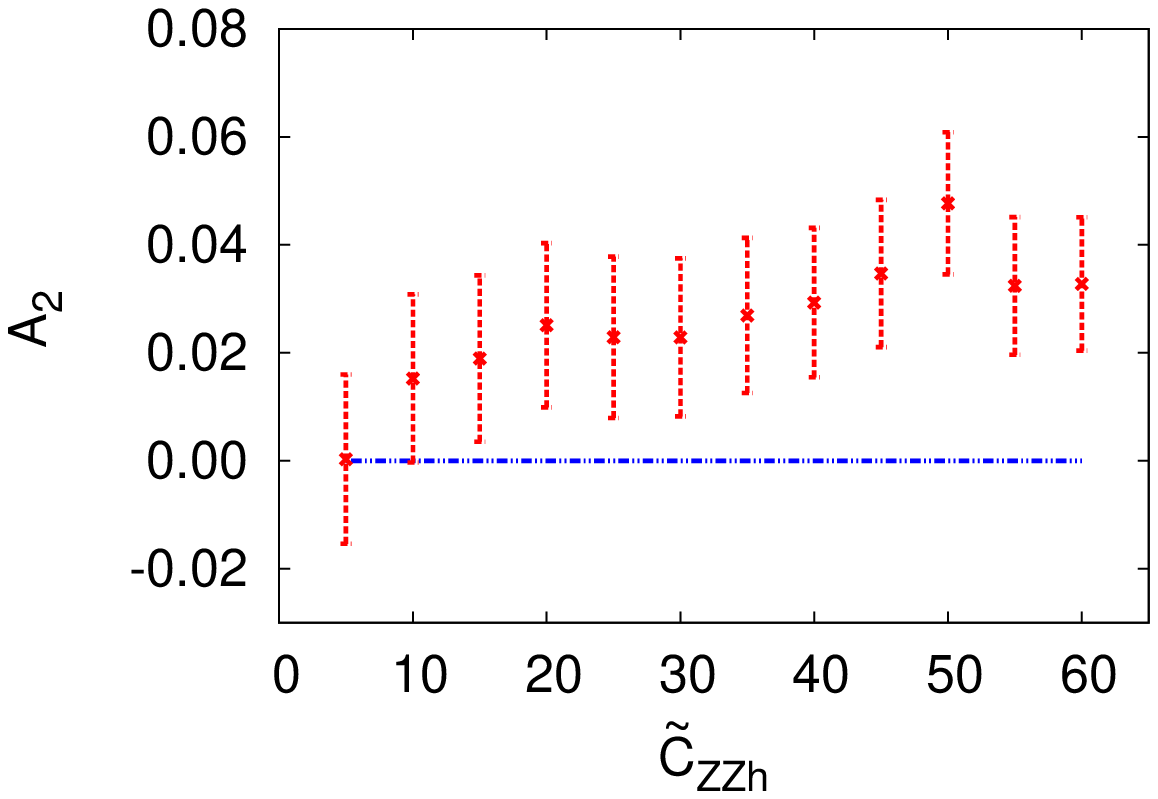}
     \caption{$\mathcal{L}= 3~{\rm ab}^{-1}$ }\label{fig:O_2_L_3000:with_gamma_zh}
   \end{subfigure}
   
    \caption{The asymmetry vs. coupling plot for the variable ${\cal O}_2$
    at Luminosities 1 ${\rm ab}^{-1}$, 1.5 ${\rm ab}^{-1}$ and 3 ${\rm ab}^{-1}$, 
   plotted for varying $\tilde{C}_{ZZh}$. The benchmark points are chosen
   such that $\tilde{C}_{\gamma Z h} = 1.5$.  }
  \label{fig:O_2_asymmetry_with_gamma_zh}
 \end{figure}

With this in mind, we plot the variation of the asymmetry against the value of the coupling
parameter $\tilde{C}_{ZZh}$ as shown in  Figures
\ref{fig:O_1_asymmetry_no_gamma_zh} to \ref{fig:O_2_asymmetry_with_gamma_zh}.
Figures \ref{fig:O_1_asymmetry_no_gamma_zh} and \ref{fig:O_2_asymmetry_no_gamma_zh}
show the variation of asymmetry for observables
${\cal O}_1$ and ${\cal O}_2$ respectively. For both of them we vary
$\tilde{C}_{ZZh}$ keeping $\tilde{C}_{\gamma Z h } = 0.0$,
wheres Figures \ref{fig:O_1_asymmetry_with_gamma_zh} and 
\ref{fig:O_2_asymmetry_with_gamma_zh} show the asymmetry plots for the same observables,
although this time we fix the $\tilde{C}_{\gamma Z h}$ coupling at 1.5. For each of the Figures
\ref{fig:O_1_asymmetry_no_gamma_zh} to \ref{fig:O_2_asymmetry_with_gamma_zh}
we present the error bars on the asymmetries for the various benchmark points. 
We have not included any specific values of systematic uncertainties which can in principle
change the asymmetries by a few percent.\\
Our calculations yield near vanishing asymmetry for SM and CP-even cases, 
which are consistent with zero within 0.5$\sigma$ for the chosen integrated luminosities.
Looking at Figures \ref{fig:O_1_asymmetry_no_gamma_zh} to \ref{fig:O_2_asymmetry_with_gamma_zh}
one can 
infer that the statistical significance improves as we go from a luminosity
reach of $1~ \rm ab^{-1}$ to $3~ \rm ab^{-1}$. 
For example, in Fig. \ref{fig:O_1_asymmetry_no_gamma_zh}, we see that for
$\tilde{C}_{ZZh} = 40.0$, the sensitivity improves from $2\sigma$ to about $4\sigma$ 
in going from $1~ \rm ab^{-1}$ to $3~ \rm ab^{-1}$.
Another important feature to notice is that the asymmetry does not
monotonically increase as we increase the magnitude of the coupling parameter. 
This is due to the fact that for higher values of couplings,
the higher order terms, {\it i.e.} beyond the interference term linear in $\tilde{C}_{ZZh}$, 
become influential. This is corroborated by the marginal decrease in the asymmetry
curve beyond  $\tilde{C}_{ZZh} = 40.0$ in Figures  \ref{fig:O_1_asymmetry_no_gamma_zh} 
and \ref{fig:O_2_asymmetry_no_gamma_zh}. 
The analysis done here agrees at the parton level to the work presented in \cite{Christensen:2010pf}, 
but once we fold in the ISR/FSR and detector effects, the statistical significance gets lowered in comparison
to what one gets for the parton level analysis.

 \section{Summary and Conclusions}
We have looked for experimentally realizing the effects of CP-violating operators
that arise from gauge-invariant dimension-6 $VVh$ couplings. Values of the couplings
consistent with electric dipole moments as well as global fits of the LHC data are taken.
We have studied {\it higgstrahlung} process ($pp \to Zh$) in presence of both CP-even and CP-odd 
couplings.
We have included effects of parton showers and also factored in detector simulation, to
make our estimates as realistic as possible.
 The presence of non vanishing asymmetries over and above the
 statistical errors is seen as the litmus test for testing the presence of CP-violating 
 new physics. To this end, we have shown that for SM and CP-conserving dimension-6 operators,
  observables which are both CP and $\hat T$ odd do not yield any asymmetry in their kinematic distributions.
 But the presence of CP-violating gauge-Higgs operators is seen to give rise to non-vanishing
 asymmetries for such observables, thus clearly demarcating the presence of CP-violating high
 scale physics. We have been able to show that for a benchmark value of $\tilde{C}_{ZZh}=40.0$, asymmetry
 can be established at about $4\sigma$ level with 3 $\rm ab ^{-1}$ in the 14 TeV run of the LHC. \\
 
 \section{Acknowledgements}
We thank Shankha Banerjee, Jyotiranjan Beuria, Tanumoy Mandal, Subhadeep Mondal, Pratishruti Saha and Juhi Dutta 
for fruitful discussions. The work of SD, BM and AS
is partially supported by funding available from the Department of Atomic Energy, Government of
India, for the Regional Centre for Accelerator-based Particle Physics (RECAPP), Harish-Chandra
Research Institute. SD acknowledges the hospitality of Indian Association for the Cultivation of
Science, Kolkata. 
%
%


\newpage
\begin{thebibliography}{99}



\bibitem{Aad:2012tfa} 
  G.~Aad {\it et al.}  [ATLAS Collaboration],
  Phys.\ Lett.\ B {\bf 716}, 1 (2012)
  [arXiv:1207.7214 [hep-ex]].
  

\bibitem{Chatrchyan:2012ufa} 
  S.~Chatrchyan {\it et al.}  [CMS Collaboration],
  Phys.\ Lett.\ B {\bf 716}, 30 (2012)
  [arXiv:1207.7235 [hep-ex]].




\bibitem{Gao:2010qx} 
  Y.~Gao, A.~V.~Gritsan, Z.~Guo, K.~Melnikov, M.~Schulze and N.~V.~Tran,
  Phys.\ Rev.\ D {\bf 81}, 075022 (2010)
  doi:10.1103/PhysRevD.81.075022
  [arXiv:1001.3396 [hep-ph]].
  



\bibitem{Choi:2002jk} 
  S.~Y.~Choi, D.~J.~Miller, M.~M.~Muhlleitner and P.~M.~Zerwas,
  Phys.\ Lett.\ B {\bf 553}, 61 (2003)
  doi:10.1016/S0370-2693(02)03191-X
  [hep-ph/0210077].
  
  


\bibitem{DeRujula:2010ys} 
  A.~De Rujula, J.~Lykken, M.~Pierini, C.~Rogan and M.~Spiropulu,
  Phys.\ Rev.\ D {\bf 82}, 013003 (2010)
  doi:10.1103/PhysRevD.82.013003
  [arXiv:1001.5300 [hep-ph]].




\bibitem{Boughezal:2012tz} 
  R.~Boughezal, T.~J.~LeCompte and F.~Petriello,
  arXiv:1208.4311 [hep-ph].

  


\bibitem{Stolarski:2012ps} 
  D.~Stolarski and R.~Vega-Morales,
  Phys.\ Rev.\ D {\bf 86}, 117504 (2012)
  doi:10.1103/PhysRevD.86.117504
  [arXiv:1208.4840 [hep-ph]].


  

\bibitem{Djouadi:2013yb} 
  A.~Djouadi, R.~M.~Godbole, B.~Mellado and K.~Mohan,
  Phys.\ Lett.\ B {\bf 723}, 307 (2013)
  doi:10.1016/j.physletb.2013.04.060
  [arXiv:1301.4965 [hep-ph]].

  
  

\bibitem{Godbole:2013lna} 
  R.~Godbole, D.~J.~Miller, K.~Mohan and C.~D.~White,
  Phys.\ Lett.\ B {\bf 730}, 275 (2014)
  doi:10.1016/j.physletb.2014.01.069
  [arXiv:1306.2573 [hep-ph]].

  


\bibitem{Ellis:2013ywa} 
  J.~Ellis, V.~Sanz and T.~You,
  Eur.\ Phys.\ J.\ C {\bf 73}, 2507 (2013)
  doi:10.1140/epjc/s10052-013-2507-6
  [arXiv:1303.0208 [hep-ph]].

  

\bibitem{Sun:2013yra} 
  Y.~Sun, X.~F.~Wang and D.~N.~Gao,
  Int.\ J.\ Mod.\ Phys.\ A {\bf 29}, 1450086 (2014)
  doi:10.1142/S0217751X14500869
  [arXiv:1309.4171 [hep-ph]].




\bibitem{Anderson:2013afp} 
  I.~Anderson {\it et al.},
  Phys.\ Rev.\ D {\bf 89}, no. 3, 035007 (2014)
  doi:10.1103/PhysRevD.89.035007
  [arXiv:1309.4819 [hep-ph]].

  

\bibitem{Delaunay:2013npa} 
  C.~Delaunay, G.~Perez, H.~de Sandes and W.~Skiba,
  Phys.\ Rev.\ D {\bf 89}, no. 3, 035004 (2014)
  doi:10.1103/PhysRevD.89.035004
  [arXiv:1308.4930 [hep-ph]].
  
  

\bibitem{Maltoni:2013sma} 
  F.~Maltoni, K.~Mawatari and M.~Zaro,
  Eur.\ Phys.\ J.\ C {\bf 74}, no. 1, 2710 (2014)
  doi:10.1140/epjc/s10052-013-2710-5
  [arXiv:1311.1829 [hep-ph]].

  

\bibitem{Gavela:2014vra} 
  M.~B.~Gavela, J.~Gonzalez-Fraile, M.~C.~Gonzalez-Garcia, L.~Merlo, S.~Rigolin and J.~Yepes,
  JHEP {\bf 1410}, 44 (2014)
  doi:10.1007/JHEP10(2014)044
  [arXiv:1406.6367 [hep-ph]].

  

\bibitem{Banerjee:2013apa} 
  S.~Banerjee, S.~Mukhopadhyay and B.~Mukhopadhyaya,
  Phys.\ Rev.\ D {\bf 89}, no. 5, 053010 (2014)
  doi:10.1103/PhysRevD.89.053010
  [arXiv:1308.4860 [hep-ph]].

  

\bibitem{Amar:2014fpa} 
  G.~Amar, S.~Banerjee, S.~von Buddenbrock, A.~S.~Cornell, T.~Mandal, B.~Mellado and B.~Mukhopadhyaya,
  JHEP {\bf 1502}, 128 (2015)
  doi:10.1007/JHEP02(2015)128
  [arXiv:1405.3957 [hep-ph]].

  

\bibitem{Ellis:2014dva} 
  J.~Ellis, V.~Sanz and T.~You,
  JHEP {\bf 1407}, 036 (2014)
  doi:10.1007/JHEP07(2014)036
  [arXiv:1404.3667 [hep-ph]].

  

\bibitem{Desai:2011yj} 
  N.~Desai, D.~K.~Ghosh and B.~Mukhopadhyaya,
  Phys.\ Rev.\ D {\bf 83}, 113004 (2011)
  [arXiv:1104.3327 [hep-ph]].
  

  

\bibitem{Bolognesi:2012mm} 
  S.~Bolognesi, Y.~Gao, A.~V.~Gritsan, K.~Melnikov, M.~Schulze, N.~V.~Tran and A.~Whitbeck,
  Phys.\ Rev.\ D {\bf 86}, 095031 (2012)
  doi:10.1103/PhysRevD.86.095031
  [arXiv:1208.4018 [hep-ph]].

  

\bibitem{Chen:2014ona} 
  Y.~Chen, A.~Falkowski, I.~Low and R.~Vega-Morales,
  Phys.\ Rev.\ D {\bf 90}, no. 11, 113006 (2014)
  doi:10.1103/PhysRevD.90.113006
  [arXiv:1405.6723 [hep-ph]].

  

\bibitem{CMS_report_h_4l_2014}
{\bf CMS Collaboration} Collaboration, {\it Constraints on anomalous
HVV interactions using H to 4l decays}, Tech. Rep. CMS-PAS-HIG-14-014,
CERN, Geneva, 2014.






\bibitem{Khachatryan:2014kca} 
  V.~Khachatryan {\it et al.} [CMS Collaboration],
  Phys.\ Rev.\ D {\bf 92}, no. 1, 012004 (2015)
  doi:10.1103/PhysRevD.92.012004
  [arXiv:1411.3441 [hep-ex]].
  
  


\bibitem{CMS_report_WWh_2014}
{\bf CMS Collaboration} Collaboration, {\it Constraints on Anomalous HWW
Interactions using Higgs boson decays to $W^+ W^-$ in the fully leptonic
final state}, Tech. Rep. CMS-PAS-HIG-14-012, CERN, Geneva, 2014.




\bibitem{Plehn:2001nj} 
  T.~Plehn, D.~L.~Rainwater and D.~Zeppenfeld,
  Phys.\ Rev.\ Lett.\  {\bf 88}, 051801 (2002)
  doi:10.1103/PhysRevLett.88.051801
  [hep-ph/0105325].

  


\bibitem{Hankele:2006ma} 
  V.~Hankele, G.~Klamke, D.~Zeppenfeld and T.~Figy,
  Phys.\ Rev.\ D {\bf 74}, 095001 (2006)
  doi:10.1103/PhysRevD.74.095001
  [hep-ph/0609075].

  
  

\bibitem{Miller:2001bi} 
  D.~J.~Miller, S.~Y.~Choi, B.~Eberle, M.~M.~Muhlleitner and P.~M.~Zerwas,
  Phys.\ Lett.\ B {\bf 505}, 149 (2001)
  doi:10.1016/S0370-2693(01)00317-3
  [hep-ph/0102023].




\bibitem{Han:2000mi} 
  T.~Han and J.~Jiang,
  Phys.\ Rev.\ D {\bf 63}, 096007 (2001)
  [hep-ph/0011271].
  
  


\bibitem{Biswal:2008tg} 
  S.~S.~Biswal, D.~Choudhury, R.~M.~Godbole and Mamta,
  Phys.\ Rev.\ D {\bf 79}, 035012 (2009)
  doi:10.1103/PhysRevD.79.035012
  [arXiv:0809.0202 [hep-ph]].

  
  

\bibitem{Hagiwara:2000tk} 
  K.~Hagiwara, S.~Ishihara, J.~Kamoshita and B.~A.~Kniehl,
  Eur.\ Phys.\ J.\ C {\bf 14}, 457 (2000)
  doi:10.1007/s100520000366
  [hep-ph/0002043].



  
\bibitem{Chang:1993jy} 
  D.~Chang, W.~Y.~Keung and I.~Phillips,
  Phys.\ Rev.\ D {\bf 48}, 3225 (1993)
  doi:10.1103/PhysRevD.48.3225
  [hep-ph/9303226].
  



\bibitem{Biswal:2009ar} 
  S.~S.~Biswal and R.~M.~Godbole,
  Phys.\ Lett.\ B {\bf 680}, 81 (2009)
  doi:10.1016/j.physletb.2009.08.014
  [arXiv:0906.5471 [hep-ph]].

  
  

\bibitem{Dutta:2008bh} 
  S.~Dutta, K.~Hagiwara and Y.~Matsumoto,
  Phys.\ Rev.\ D {\bf 78}, 115016 (2008)
  doi:10.1103/PhysRevD.78.115016
  [arXiv:0808.0477 [hep-ph]].

  
  

\bibitem{Biswal:2012mp} 
  S.~S.~Biswal, R.~M.~Godbole, B.~Mellado and S.~Raychaudhuri,
  Phys.\ Rev.\ Lett.\  {\bf 109}, 261801 (2012)
  doi:10.1103/PhysRevLett.109.261801
  [arXiv:1203.6285 [hep-ph]].

  


\bibitem{Han:2009ra} 
  T.~Han and Y.~Li,
  Phys.\ Lett.\ B {\bf 683}, 278 (2010)
  [arXiv:0911.2933 [hep-ph]].
  
  
  

\bibitem{Christensen:2010pf} 
  N.~D.~Christensen, T.~Han and Y.~Li,
  Phys.\ Lett.\ B {\bf 693}, 28 (2010)
  [arXiv:1005.5393 [hep-ph]].
  



\bibitem{Englert:2012xt} 
  C.~Englert, D.~Goncalves-Netto, K.~Mawatari and T.~Plehn,
  JHEP {\bf 1301}, 148 (2013)
  doi:10.1007/JHEP01(2013)148
  [arXiv:1212.0843 [hep-ph]].

  


\bibitem{Biswal:2005fh} 
  S.~S.~Biswal, R.~M.~Godbole, R.~K.~Singh and D.~Choudhury,
  Phys.\ Rev.\ D {\bf 73}, 035001 (2006)
  [Phys.\ Rev.\ D {\bf 74}, 039904 (2006)]
  doi:10.1103/PhysRevD.74.039904, 10.1103/PhysRevD.73.035001
  [hep-ph/0509070].


  

\bibitem{Godbole:2007cn} 
  R.~M.~Godbole, D.~J.~Miller and M.~M.~Muhlleitner,
  JHEP {\bf 0712}, 031 (2007)
  doi:10.1088/1126-6708/2007/12/031
  [arXiv:0708.0458 [hep-ph]].

  

\bibitem{Bhattacherjee:2015xra} 
  B.~Bhattacherjee, T.~Modak, S.~K.~Patra and R.~Sinha,
  arXiv:1503.08924 [hep-ph].
  
  
  

\bibitem{Beneke:2014sba} 
  M.~Beneke, D.~Boito and Y.~M.~Wang,
  JHEP {\bf 1411}, 028 (2014)
  [arXiv:1406.1361 [hep-ph]].
  



\bibitem{Godbole:2014cfa} 
  R.~M.~Godbole, D.~J.~Miller, K.~A.~Mohan and C.~D.~White,
  JHEP {\bf 1504}, 103 (2015)
  [arXiv:1409.5449 [hep-ph]].
  
  
  
 

\bibitem{Dawson:2013owa} 
  S.~Dawson, S.~K.~Gupta and G.~Valencia,
  Phys.\ Rev.\ D {\bf 88}, no. 3, 035008 (2013)
  [arXiv:1304.3514 [hep-ph]].
  
  

\bibitem{Kruse:2014pya} 
  A.~Kruse, A.~S.~Cornell, M.~Kumar, B.~Mellado and X.~Ruan,
  Phys.\ Rev.\ D {\bf 91}, no. 5, 053009 (2015)
  doi:10.1103/PhysRevD.91.053009
  [arXiv:1412.4710 [hep-ph]].


  

\bibitem{Rattazzi:1988ye} 
  R.~Rattazzi,
  Z.\ Phys.\ C {\bf 40}, 605 (1988).
  doi:10.1007/BF01560232
  

  

\bibitem{Hagiwara:1993sw} 
  K.~Hagiwara and M.~L.~Stong,
  Z.\ Phys.\ C {\bf 62}, 99 (1994)
  doi:10.1007/BF01559529
  [hep-ph/9309248].
  
  


\bibitem{Gounaris:1995mx} 
  G.~J.~Gounaris, F.~M.~Renard and N.~D.~Vlachos,
  Nucl.\ Phys.\ B {\bf 459}, 51 (1996)
  doi:10.1016/0550-3213(95)00602-8
  [hep-ph/9509316].
  
  
  

\bibitem{Niezurawski:2004ga} 
  P.~Niezurawski, A.~F.~Zarnecki and M.~Krawczyk,
  Acta Phys.\ Polon.\ B {\bf 36}, 833 (2005)
  [hep-ph/0410291].
  
  
  

\bibitem{Skjold:1995jp} 
  A.~Skjold and P.~Osland,
  Nucl.\ Phys.\ B {\bf 453}, 3 (1995)
  doi:10.1016/0550-3213(95)00373-Z
  [hep-ph/9502283].

  



\bibitem{Skjold:1993jd} 
  A.~Skjold and P.~Osland,
  Phys.\ Lett.\ B {\bf 311}, 261 (1993)
  doi:10.1016/0370-2693(93)90565-Y
  [hep-ph/9303294].



  
\bibitem{Skjold:1994qn} 
  A.~Skjold and P.~Osland,
  Phys.\ Lett.\ B {\bf 329}, 305 (1994)
  doi:10.1016/0370-2693(94)90777-3
  [hep-ph/9402358].
  
  
  
  
\bibitem{Grzadkowski:1995rx} 
  B.~Grzadkowski and J.~F.~Gunion,
  Phys.\ Lett.\ B {\bf 350}, 218 (1995)
  doi:10.1016/0370-2693(95)00369-V
  [hep-ph/9501339].
  
  
  
  

\bibitem{Godbole:2006uy} 
  R.~M.~Godbole,
  Pramana {\bf 67}, 835 (2006).
  doi:10.1007/s12043-006-0096-8

  

  
\bibitem{Zhang:2003it} 
  B.~Zhang, Y.~P.~Kuang, H.~J.~He and C.~P.~Yuan,
  Phys.\ Rev.\ D {\bf 67}, 114024 (2003)
  doi:10.1103/PhysRevD.67.114024
  [hep-ph/0303048].
  
  
  


\bibitem{Buszello:2006hf} 
  C.~P.~Buszello and P.~Marquard,
  hep-ph/0603209.
  
  
  

\bibitem{Hankele:2006ja} 
  V.~Hankele, G.~Klamke and D.~Zeppenfeld,
  hep-ph/0605117.
  
  
  
  
%


  
  
  
  
  


\bibitem{Buchmuller:1985jz} 
  W.~Buchmuller and D.~Wyler,
  Nucl.\ Phys.\ B {\bf 268}, 621 (1986). 
  
  
  
\bibitem{Grzadkowski:2010es} 
  B.~Grzadkowski, M.~Iskrzynski, M.~Misiak and J.~Rosiek,
  JHEP {\bf 1010}, 085 (2010)
  [arXiv:1008.4884 [hep-ph]].
    
  


\bibitem{Contino:2013kra} 
  R.~Contino, M.~Ghezzi, C.~Grojean, M.~Muhlleitner and M.~Spira,
  JHEP {\bf 1307}, 035 (2013)
  doi:10.1007/JHEP07(2013)035
  [arXiv:1303.3876 [hep-ph]].
  
 

\bibitem{Leung:1984ni} 
  C.~N.~Leung, S.~T.~Love and S.~Rao,
  Z.\ Phys.\ C {\bf 31}, 433 (1986).
  doi:10.1007/BF01588041
  
  

  
\bibitem{Hagiwara:1993qt} 
  K.~Hagiwara, R.~Szalapski and D.~Zeppenfeld,
  Phys.\ Lett.\ B {\bf 318}, 155 (1993)
  doi:10.1016/0370-2693(93)91799-S
  [hep-ph/9308347].
  
  
  

\bibitem{GonzalezGarcia:1999fq} 
  M.~C.~Gonzalez-Garcia,
  Int.\ J.\ Mod.\ Phys.\ A {\bf 14}, 3121 (1999)
  [hep-ph/9902321].
  
  

 

\bibitem{Chien:2015xha} 
  Y.~T.~Chien, V.~Cirigliano, W.~Dekens, J.~de Vries and E.~Mereghetti,
  JHEP {\bf 1602}, 011 (2016)
  [JHEP {\bf 1602}, 011 (2016)]
  doi:10.1007/JHEP02(2016)011
  [arXiv:1510.00725 [hep-ph]].
  
  
  
\bibitem{Cirigliano:2016njn} 
  V.~Cirigliano, W.~Dekens, J.~de Vries and E.~Mereghetti,
  arXiv:1603.03049 [hep-ph].
  
  
  
  
  
\bibitem{Dwivedi:2015nta} 
  S.~Dwivedi, D.~K.~Ghosh, B.~Mukhopadhyaya and A.~Shivaji,
  Phys.\ Rev.\ D {\bf 92}, no. 9, 095015 (2015)
  doi:10.1103/PhysRevD.92.095015
  [arXiv:1505.05844 [hep-ph]].
  
  
  
  

  
\bibitem{Choudhury:2012tk} 
  D.~Choudhury, R.~Islam and A.~Kundu,
  Phys.\ Rev.\ D {\bf 88}, no. 1, 013014 (2013)
  doi:10.1103/PhysRevD.88.013014
  [arXiv:1212.4652 [hep-ph]].
  
  
  
  
  
\bibitem{Dahiya:2013uba} 
  M.~Dahiya, S.~Dutta and R.~Islam,
  Phys.\ Rev.\ D {\bf 93}, no. 5, 055013 (2016)
  doi:10.1103/PhysRevD.93.055013
  [arXiv:1311.4523 [hep-ph]].
   
  
  
 
\bibitem{Valencia:1994zi} 
  G.~Valencia,
  In *Boulder 1994, Proceedings, CP violation and the limits of the standard model* 235-269, and Iowa State U. Ames - AMES-HET-94-12 (94,rec.Dec.) 35 p
  [hep-ph/9411441].
     
  
  
  
%
%
  
  

   
  



  \bibitem{Alwall:2014hca} 
  J.~Alwall, R.~Frederix, S.~Frixione, V.~Hirschi, F.~Maltoni, O.~Mattelaer, H.-S.~Shao and T.~Stelzer {\it et al.},
  JHEP {\bf 1407}, 079 (2014)
  [arXiv:1405.0301 [hep-ph]].
  
  
 
  
\bibitem{Christensen:2008py} 
  N.~D.~Christensen and C.~Duhr,
  Comput.\ Phys.\ Commun.\  {\bf 180}, 1614 (2009)
  [arXiv:0806.4194 [hep-ph]].
   
   
   
   
\bibitem{Ball:2012cx} 
  R.~D.~Ball {\it et al.},
  Nucl.\ Phys.\ B {\bf 867}, 244 (2013)
  doi:10.1016/j.nuclphysb.2012.10.003
  [arXiv:1207.1303 [hep-ph]].
   
   
   
   
   
\bibitem{Sjostrand:2006za} 
  T.~Sjostrand, S.~Mrenna and P.~Z.~Skands,
  JHEP {\bf 0605}, 026 (2006)
  doi:10.1088/1126-6708/2006/05/026
  [hep-ph/0603175].
   


\bibitem{deFavereau:2013fsa} 
  J.~de Favereau {\it et al.} [DELPHES 3 Collaboration],
  JHEP {\bf 1402}, 057 (2014)
  doi:10.1007/JHEP02(2014)057
  [arXiv:1307.6346 [hep-ex]].
  

  
  
  \bibitem{root_package}  
https://root.cern.ch/
  
  
 
 
\bibitem{Banerjee:2015bla} 
  S.~Banerjee, T.~Mandal, B.~Mellado and B.~Mukhopadhyaya,
  JHEP {\bf 1509}, 057 (2015)
  [arXiv:1505.00226 [hep-ph]].
  
  
  
  
%
%
%
  
  
%

  
\end{thebibliography}
\end{document}